\documentclass{aa}
\usepackage{graphicx}
\usepackage{epsfig}
\usepackage{longtable,lscape}
\usepackage{natbib}
\usepackage{rotating}
\usepackage{amssymb}
\bibpunct{(}{)}{;}{a}{}{,} 
\begin{document}

\title{An optical/NIR survey of globular clusters in early-type galaxies}
\subtitle{I. Introduction and data reduction procedures\thanks{Table 4 is only available in electronic form
at the CDS via anonymous ftp to cdsarc.u-strasbg.fr (130.79.128.5)
or via http://cdsweb.u-strasbg.fr/cgi-bin/qcat?J/A+A/}}

\author{A. L. Chies-Santos\inst{1}, S. S. Larsen\inst{1}, E. M. Wehner\inst{1, }\inst{2}, H. Kuntschner\inst{3}, J. Strader\inst{4} and J. P. Brodie\inst{5}}

\offprints{A.L.Chies@uu.nl} 

\institute{Astronomical Institute, University of Utrecht, Princetonplein 5, NL-3584, Utrecht, The Netherlands
\and Department of Physics and Astronomy, McMaster University, Hamilton L8S 4M1, Canada 
\and Space Telescope European Coordinating Facility, European Southern Observatory, Garching, Germany
\and Harvard-Smithsonian Center for Astrophysics, Cambridge, MA 02138, USA 
\and UCO/Lick Observatory, University of California, Santa Cruz, CA 95064, USA}

\date{Received 3 September 2010 / Accepted 1 October 2010}

\abstract
{The combination of optical and near-infrared (NIR) colours has the potential to break the age/metallicity degeneracy and offers a better metallicity sensitivity than optical colours alone. Previous tudies of extragalactic globular clusters (GCs) with this colour combination, however, have suffered from small samples or have been restricted to a few galaxies.
}
{We compile a homogeneous and representative sample of GC systems with multi-band photometry to be used in subsequent papers where ages and metallicity distributions will be studied.
}
{We acquired deep K-band images of 14 bright nearby early-type galaxies. The images were obtained with the LIRIS near-infrared spectrograph and imager at the William Herschel Telescope (WHT) and combined with optical ACS g and z images from the Hubble Space Telescope public archive.}
{For the first time GC photometry of 14 galaxies are observed and reduced homogeneously in this wavelength regime. We achieved a limiting magnitude of $K\sim 20 - 21$. For the majority of the galaxies we detect about 70 GCs each. NGC\,4486 and NGC\,4649, the cluster-richest galaxies in the sample contain 301 and 167 GCs, respectively. We present tables containing coordinates, photometry and sizes of the GCs available. 
}
{}


\titlerunning{GCS of early-type galaxies in the NIR}

\authorrunning{Chies-Santos et al.}

\maketitle

\section{Introduction}
Understanding how galaxies formed and evolved is one of the fundamental problems in modern astrophysics. Spheroidal systems (ellipicals, lenticulars and spiral bulges) contain $\sim75\%$ of the Local Universe's stellar content (\citealt{renzini06}), and much of our knowledge on galaxy formation and evolution comes from detailed studies of galaxies in this close portion of the Universe. 
In the hierarchical merging framework of structure formation, early-type galaxies are the products of mergers, and star formation is governed by the amount of gas involved in the merger event (eg. \citealt{kauf93}). This model is very successful in explaining structure formation, being the product of the $\lambda \rm CDM$ cosmology (eg. \citealt{kom09}), which is the best at describing the observations of the cosmic microwave background, large scale structure and the accelerated expansion of the Universe with the Big Bang theory. 
Nonetheless, it remains to be understood how baryonic galaxy formation occurs within this framework.

Globular clusters (GCs) are found around all major galaxies, and young massive clusters (YMCs) are being observed in many mergers and standard spirals (\citealt{ws95}; \citealt{lr99}). If (massive) galaxies (early-type) formed through mergers, these YMCs are good candidates to be the progenitors of the GCs we observe today in many early-type galaxies. This ubiquity of star clusters around all major galaxies suggests that GC formation traces important events in the history of their parent galaxies.
Even though we cannot state anymore that all GCs are simple stellar populations (SSPs) due to the evidence of multiple populations in several Galactic (eg. \citealt{bedin04}, \citealt{piotto07}) and Large Magellanic cloud star clusters (\citealt{santiago02}; \citealt{mackey08}), they are still more easily modelled than the mix of stellar populations that make up the integrated light of a galaxy. 

A thorough understanding of the evolutionary histories of galaxies must involve both their GC systems and the integrated light. Integrated light studies suffer from a degeneracy between age, metallicity and burst strength, which is difficult to overcome. GCs on the other hand, have the advantage of being discrete and separable star formation tracers.
Generally, a significant population of GCs with a particular age would be a strong indication that the host galaxy experienced a burst of star formation at the corresponding epoch. 
If such a burst occurred recently, within the last few Gyrs, and involved a significant fraction of the stars, it should also be detectable in the integrated light. 
A significant fraction, if not the vast majority of all the stars are formed in clusters (\citealt{ll03}) and a considerable part of these do not remain bound for a Hubble time and probably help build up the integrated light of a galaxy. Therefore, a galaxy's integrated light should be correlated in many ways to that of its GC system.
Studies of GCs and their host galaxies are thus complementary and evidence of star formation seen in the integrated light should also leave imprints in the GC system.

The colour distribution of GC systems in most galaxies is bi/multimodal in optical colours. Many studies, especially spectroscopic ones (eg. \citealt{strader07} for direct and \citealt{strader05}, \citealt{cenarro07} for indirect evidence), show that this feature is due to 2 old populations ($\gtrsim$\,10 Gyr) that differ in metallicity.
Bimodal colour (metallicity) distributions may indicate distinct cluster formation mechanisms for the two populations, with age differences of a few Gyrs  ($\sim 2$ Gyrs) allowed within the uncertainties. Moreover there is sufficient evidence available (\citealt{bs06}) to place metal-rich GC formation along with the bulk of the field stars in their parent galaxies.
Most GC formation scenarios introduced to account for colour bimodality (\citealt{az92}, \citealt{fbg97}, \citealt{cote98}, \citealt{beasley02}, \citealt{strader05} and \citealt{rhode05}) assume different formation/assembly channels for the red and blue clusters although these clusters have very similar internal properties such as masses and sizes. \cite{muratov10} propose a common mechanism for the formation of both sub-populations in which the bimodal metallicity distribution arises naturally in many of the model realisations from the combination of the history of galaxy assembly and the amount of cold gas in protogalactic systems. This scenario is built upon the hierarchical merging framework, in which early mergers of smaller hosts create only blue clusters, and subsequent mergers of progenitor galaxies produce both blue and red clusters.
\cite{yoon06}, however, challenged the metallicity bimodality interpretation. These authors claim that the horizontal branch stars (HB) bring up a non-linear behavior on the metallicity-colour relation of GCs. This non-linearity is in turn responsible for making a coeval group of clusters with a unimodal metallicity spread exhibit colour bimodality. 

We want to understand metallicity distributions of GC systems and how their age distributions are shaped. The most direct way to measure the metallicity and derive ages for GC systems would be to take spectra of its individual members. Spectroscopy for a large number of GCs is very time consuming; it is difficult to collect data for representative samples in many galaxies, although there is an ongoing effort of the extragalactic GC community (\citealt{cohen98}, \citealt{strader05}, \citealt{puzia05}, \citealt{cenarro07}) to collect a significant amount of data for a large number of galaxies.
The majority of spectroscopic studies reach around 2 dozen clusters (\citealt{strader05}, \citealt{puzia05}) biased to the brightest parts. It still remains unclear if these are representative of the system as a whole, with the exception of a few cases (\citealt{puzia05a} on M\,31 reached $\sim$70 GCs; \citealt{woodley09} on NGC\,5128 and \citealt{cohen98} on M\,87 who reached over 100 clusters).

The general result is that the studied clusters are predominantly old ($\gtrsim 10$ Gyrs) but there is some disagreement on the fraction of GCs with ages of $5 - 6$ Gyrs.
While some studies (eg. \citealt{strader05}, \citealt{cohen98}) find little evidence for young GCs, the study by \cite{puzia05} finds that up to 1/3 of their 143 clusters in different E/S0 galaxies are younger than $10$ Gyrs.
Due to small number statistics, metallicity distributions of GC systems in early type galaxies are poorly constrained spectroscopically. M\,87 (\citealt{cohen98}) shows marginal evidence for a bimodal as opposed to unimodal metallicity distribution. Nevertheless, there is evidence that this could also be due to the \textit{blue tilt} or luminosity-colour relation (see eg. \citealt{strader06}, \citealt{harris06}) as \cite{strader07} reanalysed the \cite{cohen03} sample of GCs in NGC\,4472, which does not show the \textit{blue tilt} (\citealt{strader06}) and found bimodal metallicity distributions. 

Given the difficulty of current spectroscopic studies in reaching representative statistics, NIR imaging appears as a powerful alternative. One can achieve much better statistics, reaching fainter GCs than the deepest spectroscopic studies with shorter integration times. It has been shown that the colours that best represent the true metallicity distributions are the combination of optical and NIR (eg. \citealt{puzia02}; \citealt{cb07}). While the NIR depends mainly on the red giant branch being metallicity sensitive, the optical depends on stars near the main sequence turn-off point being both metallicity and age sensitive. By using the optical/NIR combination we have, therefore, a better chance to break the age-metallicity degeneracy inherent in optical colours alone. Hence, studying GCs in the NIR is an alternative to spectroscopy in revealing their actual metallicity distributions.

There have been several attempts to study GC systems with NIR imaging with a variety of instruments (\citealt{hempel07}, \citealt{hempel07AA}, \citealt{kotulla08} and \citealt{bs06} for a compilation).
With the exception of a few cases, the detectors had generally small fields of view ($\sim\,2 \times 2 \,arcmin^2$). In total around a dozen systems have been studied with the optical/NIR technique.
 The general result is that most of the studied galaxies host predominantly old clusters, but there is still a possible significant fraction of intermediate age GCs in some, with the strongest and most extensively studied cases being NGC\,1316, NGC\,4365 and NGC\,5846. While the first one has a power-law rather than log-normal GC luminosity function for the metal rich GCs (\citealt{goud01b}), which is strong evidence for a younger population, the GC systems of the other two do not present any outstanding feature. In turn, the central light of NGC\,4365 is found to be uniformly old (\citealt{davies01}; \citealt{yamada06}). 
As for metallicity distributions using the optical/NIR approach there has been only one case so far. \cite{kz07} studied the (I-H) colour distribution for 80 GCs in M\,87 with NICMOS and WFPC2 data and found a clear bimodal distribution. 

Improvements in NIR SSP models over the past few years have been significant (eg. \citealt{m05} and \citealt{marigo08} for the TP-AGB, \citealt{percival09} for $\alpha - $enhancement estimates). Along with this progress, the full potential of the optical/NIR technique should be further explored. 
The purpose of this project is to establish a \textit{large} and \textit{homogeneous} optical/near-infrared data set of GC systems in elliptical and lenticular galaxies where differential comparisons among the different GC systems are possible.
In the present paper the survey is introduced, the data reduction procedures are thoroughly explained and catalogues of the GC systems with $g$, $z$ and $K$ photometry and sizes are made available.
Ages and metallicity distributions will be studied in subsequent papers.
\section{Galaxy sample selection}

The target galaxies were selected from the SAURON sample \citep{zeeuw02}. Of the 72 galaxies that make up the SAURON survey, we added three more cases that were observed with the instrument and restricted ourselves to 16 of these 75 galaxies that were observed with HST/ACS in g (F475W) and z (F850LP) to keep the sample homogeneous. Of these 16, we obtained K-band imaging for 14. Two of the 16 were not observed due to bad weather conditions and override programs during the observing runs.
The SAURON data provide a wide range of
kinematic and stellar population parameters for the galaxies, including line indices sensitive
to age and metallicity (H$\beta$, Fe5015, Mg$b$) and their gradients, presence of distinct kinematic
components and rotation parameters. 
We required distance moduli $(m-M) < 32$ 
and absolute magnitudes $M_B < -19$ to ensure the detection of significant numbers of GCs. 
While this sample is not representative of early-type galaxies as a whole because we do not cover low-mass galaxies ($M_B >-19$), it is quite representative within the SAURON sample, including galaxies with either boxy or disky isophotes, being slow or fast rotators, and having a variety of kinematic properties. The characteristics of these galaxies are listed in Tables \ref{sample} and \ref{sauron} where we also list estimates of ages and metallicities from integrated light studies (\citealt{sb06}, \citealt{yamada06}, \citealt{kunt10}) and the estimated number of GCs and their specific frequency ($S_N$) (\citealt{peng08} and \citealt{kw01}).  

Our sample contains 10 galaxies classified as ellipticals and 4 as S0's; 5 are slow rotators and 9 are fast rotators. All but two of the galaxies (NGC\,3377 and NGC\,4278) belong to the Virgo Cluster of Galaxies.
In the following a short description of earlier work on the 14 GC systems is presented, and the most important results on kinematics and integrated stellar populations from the SAURON survey are summarised.

\textbf{NGC\,3377:\,}The GC system of this E5/E6 type galaxy, located in Leo Group I, was studied by \cite{kw01} with (V-I), and a likely bimodal colour distribution was found.
It was suggested by \cite{kormendy98} that this galaxy contains a black hole.
Contains spatially extended\footnote{spatially extended, i.e covering large parts of the observed field of view (\citealt{kunt10}).} signs of SSP equivalent \footnote{SSP equivalent ages, ie., when any observable stellar population parameter is the one the object would have if it formed at a single age with a single chemical composition (\citealt{trager08}).} younger ages (\citealt{kunt10}).

\textbf{NGC\,4278:\,}\cite{forbes96a} studied the GC luminosity function of this E1/E2 galaxy and confirmed its location in the Coma I cloud. \cite{forbes96} studied the (V-I) colour distribution of the GC system finding no obvious bimodality. Later on, \cite{kw01} with (V-I), found a likely bimodal colour distribution. 
This galaxy hosts a LINER AGN (\citealt{veron06}). \citealt{kunt10} report ages consistent with old stellar populations.

The following galaxies are also part of the ACS Virgo Cluster Survey (ACSVCS) (\citealt{jordan04}, \citealt{peng06}), which made use of the (g-z) colour. 
Unless otherwise stated, this is the colour used in the GC systems studies.
We refer the reader to the appendix of \cite{ferrarese06} for notes on the morphology of the individual galaxies.

\textbf{NGC\,4365:\,}The GC system of this E3 galaxy was studied by \cite{forbes96} and subsequently by \cite{kw01} and \cite{larsen01} who found no obvious bimodality in the (V-I) colour distribution. \cite{peng06} reported a bimodal distribution with a broad red peak and an excess of red GCs when compared to similar luminosity ACSVCS galaxies.
This galaxy has also been studied with optical/NIR photometry by \cite{puzia02}, by \cite{lbs05} with V, I and K, by \cite{kundu05} with g, I and H and spectroscopically by \cite{larsen03} and \cite{brodie05}. Some of these authors find evidence for a significant fraction of young GCs. It hosts a kinematically distinct component (KDC) (\citealt{bender98}) and has ages consistent with old stellar populations (\citealt{davies01}). 

\textbf{NGC\,4374:\,}Also known as M\,84, the GC system of this galaxy was studied by \cite{gr04} with (B-R) and by \cite{peng06}, who find a bimodal colour distribution. This last study finds a sparsely populated, blue peak when compared to same luminosity-type ACSVCS early-type galaxies. It is a Seyfert 2 AGN (\citealt{veron06}). \cite{kunt10} report ages consistent with old stellar populations. 

\textbf{NGC\,4382:\,}\cite{peng06} report the GC system of this galaxy, also known as M\,85, to be the best case for a trimodal colour distribution among the ACSVCS-studied galaxies. It contains a spatially constrained, central (post) starburst\footnote{spatially constrained central starburst, i.e, the star formation is connected to a thin embedded disk (\citealt{kunt10}).}  (\citealt{kunt10}, \citealt{shapiro10}).

\textbf{NGC\,4406:\,} This S0/E3 galaxy, also known as M86, is a noteworthy galaxy with on-going dust stripping (\citealt{ferrarese06}). It is also part of the studies by \cite{forbes96} and \cite{larsen01}, with no strong bimodality being reported. Both studies were carried out with (V-I). Similar results were found by \cite{kw01} (with V-I photometry) and \cite{peng06}. It was also studied with B, V and R photometry, by \cite{rz04} who reported a multimodal colour distribution. It hosts a KDC (\citealt{bender98}).

\textbf{NGC\,4473:\,} While \cite{kw01} (with V-I photometry) and \cite{peng06} report a bimodal colour distribution for this system, \cite{larsen01} (with V-I photometry) find marginally significant bimodality. \cite{kunt10} report ages consistent with old stellar populations.

\textbf{NGC\,4486:\,}The Virgo central giant elliptical, also know as M87, is the most extensively studied galaxy in the sample, hosting the largest numbers of clusters ($\sim$15000, \citealt{peng08}). \cite{es96}, \cite{larsen01} both with (V-I) and \cite{peng06} report significant colour bimodality. The GC system of this galaxy has been reported to host a correlation between colour and luminosity for individual, metal-poor GCs, the blue tilt (\citealt{strader06} both with (g-z), \citealt{mieske06}, \citealt{peng09}, \citealt{harris09} with g',r' and i). Evidence for significant, intermediate-age population (\citealt{kpbm02}) is not found through optical/NIR imaging  with V, I and K photometry. \cite{kz07} find a clear optical/NIR (I-H) bimodal distribution. \cite{cohen98} present a large spectroscopic sample of GCs in M\,87, and they are inferred to be uniformly old ($>10$ Gyrs). A clear bimodality in [Fe/H] and Mgb is not seen in that study. It is a LINER AGN (\citealt{veron06}). \cite{kunt10} report ages consistent with old stellar populations.

\textbf{NGC\,4526:\,}\cite{peng06} report significant colour bimodality. Contains residual, ongoing star formation in a central disk/ring structure (\citealt{kunt10}). 
\cite{shapiro10} estimate the star formation rate for this galaxy to be $0.37\,M_{\sun}\,$yr$^{-1}$. 

\textbf{NGC\,4552:\,}The GC system of this galaxy, also known as M\,89, was studied by \cite{kw01} and \cite{larsen01} with (V-I) colour and by \cite{peng06} who report likely colour bimodality. It is a Seyfert 2 AGN (\citealt{veron06}) and hosts a KDC (\citealt{emsellem07}). A pronounced $H\beta$ absorption turn-up is seen in this galaxy, although the presence of a spatially constrained, central (post) starburst is uncertain (\citealt{kunt10}).

\textbf{NGC\,4570:\,} Like NGC\,4374 the colour distribution of this galaxy is likely bimodal (\citealt{peng06}), with a small blue peak if compared to similar luminosity-type ACSVCS early-type galaxies. \cite{kotulla08} find a significant population of intermediate-age globular clusters with g, z and K optical/NIR photometry. \cite{kunt10} report ages consistent with old stellar populations. 

\textbf{NGC\,4621:\,}\cite{kw01} with (V-I) colour and \cite{peng06} report obvious bimodality. This galaxy hosts a KDC (\citealt{emsellem07}) and \cite{kunt10} report ages consistent with old stellar populations. 

\textbf{NGC\,4649:\,}\cite{kw01} and \cite{larsen01} with (V-I) and \cite{peng06} report significant colour bimodality. 
\cite{forbes04} find that the underlying galaxy starlight has a similar density profile slope and (g-i) colour to the red GCs. The GC system of this galaxy, also know as M60, has been reported to exhibit the blue tilt (\citealt{strader06}, \citealt{mieske06}, \citealt{lee08}). Some GCs have also been observed spectroscopically (\citealt{bridges06}, \citealt{pierce06}, \citealt{hwang08}). \cite{bridges06} concluded that the GC system exhibits no rotation, and a dark-matter halo is inferred from its kinematics. \cite{pierce06} find most GCs to be old ($>10$\,Gyr), but several intermediate-ages are present ($2-3$ Gyrs). These authors also find a trend of decreasing $\alpha$-element ratio with increasing metallicity. \cite{hwang08}, on the other hand, report that the NGC4649 GC system exhibits significant overall rotation and also a dark-matter halo. It hosts a black hole (\citealt{gebhardt03}).

\textbf{NGC\,4660:\,}The GC system of this galaxy is found to be not significantly bimodal (\citealt{kw01} with the (V-I) colour, \citealt{peng06}). \cite{kunt10} report ages consistent with old stellar populations.   

It appears that some studies find colour bimodality while others do not, even for the same galaxies (eg. NGC\,4278).
These differences can be attributed to the different colours used in the different studies, which sample stars in different evolutionary stages.
Another factor for different shapes in the colour distributions are the different data sets used in different studies. Different data sets have different photometric uncertainties with large uncertainties likely causing a bimodal distribution to become unimodal if the range in wavelength is too narrow.  

\begin{landscape}
\begin{table}
\begin{scriptsize}
\centering
\begin{tabular}{cccccccccccccc}

\hline
(1)         & (2)            & (3)       &             (4)       &      (5)         &    (6)       &        (7) &          (8)           &   (9)      &    (10)                 &         (11)   &          (12)    &    (13)    &    (14)  \\
\hline
Galaxy &$\alpha$&$\delta$& Morphological& $\rho_E$&  $M_B$  &  (m-M) &         AGN         & age(Gyr)  & [Z/H] & age(Gyr) & [Z/H] &  N$_{GCs}$ & $S_N$ \\
             &(J2000)  & (J2000)& Type & $ [Mpc^{-3}] $&  &          &    &SB+06   &SB+06+   &   Yamada06+    &  Yamada06+ &&       \\
\hline
NGC 3377 & 10 47 42.4 &+13 59 08 &E5/E6          &0.49 & $-19.2$   &			   30.25 &yes*& 8.90 & -0.10&-&-&$266\pm66$&$2.4\pm0.6$\\
NGC 4278 &12 20 06.8  &+29 16 51 &E1/E2          &1.25 & $-21.2$   &		    31.03 & LINER&11.64&0.079 &-&-&-&-\\
NGC 4365 & 12 24 28.2 &+07 19 03 &E3             &2.93 & $- 20.96$        &	31.55 & -&12.60&0.139 &14&0.16&3246$\pm$598&3.86$\pm$0.71\\
NGC 4374 &12 25 03.7  &+12 53 13 &E1             &3.99 & $-21.2$   &			  31.32 & Sy2  & 10.40&0.099 &-&-&4301$\pm$1201&5.20$\pm$1.45\\
NGC 4382 &12 25 24.2  &+18 11 27 &S0$^{+}(s)pec$ &2.04 & $-21.3$  &    31.33 & - & -&- &-&-&1110$\pm$181&1.29$\pm$0.21\\
NGC 4406 & 12 26 11.7 &+12 56 46 &S0/E3          &1.41 &$-20.60$    &						       31.17 &-&- & -&-&-&2660$\pm$129&2.57$\pm$0.12\\
NGC 4473 &12 29 48.9  &+13 25 46 &E5             &2.17 & $-20.3$  &			     30.98 & -         & -&- &12.3&0.09&376$\pm$97&1.98$\pm$0.51\\
NGC 4486 &12 30 49.4  &+12 23 28 &E0/E1$^{+}pec$ &4.17 & $-21.79$ &  31.03 & LINER     & -& -&-&-&14660$\pm$891&12.59$\pm$0.77\\
NGC 4526 &12 34 03.0  &+07 41 57 &SAB0$^{0}(s)$  &2.97 & $-20.0$  & 31.14  &   - &- & -&-&-&388$\pm$117&1.09$\pm$0.33\\
NGC 4552 &12 35 39.8  &+12 33 23 & E0/E1         &2.97 & $-20.6$    &		       30.93  &  Sy2 & 8.89 &0.121 &-&-&984$\pm$198&2.82$\pm$0.57\\
NGC 4570 &12 36 53.6  &+07 14 46 &S0 sp          &2.66 & $-19.5$  &		    30.91  & - &- &- &-&-&139$\pm$23&1.09$\pm$0.18\\
NGC 4621 &12 42 02.3  &+11 38 49 &E5             &2.60 & $-20.6$  &			  31.31  &- & 11.40 &0.115 &13.3&0.23&803$\pm$355&2.70$\pm$1.19\\
NGC 4649 &12 43 40.0 &+11 33 10 &E2              &3.49 & $-21.47$        &								31.13  & yes*  & -& -&-&-&4745$\pm$1099&5.16$\pm$1.20\\
NGC 4660 &12 44 32.1 &+11 11 25 &E               &3.37 & $-19.2$  &			    30.54  &  -& -& -&-&-&205$\pm$28&2.97$\pm$0.41\\
\hline

\end{tabular}
\caption{Galaxy sample characteristics: (1) galaxy, (2),(3) equatorial coordinates, (4) morphological type (from \citealt{emsellem07} and NED), (5) environmental density ($\rho_E$) (from Tully 1988), (6) integrated magnitude (HYPERLEDA), (7) distance modulus (\citealt{tonry01} and NED), (8) AGN class (\citealt{veron06})--*there is evidence for a black hole, although the classification of the AGN is missing (\citealt{kormendy98}, \citealt{gebhardt03}), (9,10,11,12) ages and metallicities estimates (from the spectral synthesis analysis of \citealt{sb06} and H$\beta$vs.[MgFe] estimation of \citealt{yamada06} respectively). The last two columns refer to the (13) estimated number of GCs ($N_{GCs}$) and the (14) specific frequency ($S_N$) from \cite{peng08} for the ACSVCS galaxies and \cite{kw01} for NGC\,3377.}
\label{sample}
\end {scriptsize}
\end{table}

\begin{table}
\begin{scriptsize}
\centering
\begin{tabular}{cccccccc}
\hline
(1)         & (2)            & (3)       &             (4)       &      (5)         &    (6)       &        (7) &          (8)         \\
\hline
Galaxy & $a_4$  & $\lambda_R$ & Rotator&  GRP &  Age(Gyr)  & Metallicity &  $[\alpha/Fe]$    \\
              &               &                           & (Fast/Slow)  &&Re/8 \& Re    &Re/8 \& Re   &Re/8 \& Re   \\
\hline 
NGC 3377  & 0.94      & 0.475  & F &SC      &$6.7^{+0.7}_{-0.6}$\& $7.7^{+0.7}_{-0.7}$           & $0.01\pm0.04$ \&$-0.21\pm0.04$ &      $0.24\pm0.04$ \&$0.20\pm0.04$      \\
NGC 4278  &$-0.15$ & 0.149  & F &MC     &$17.7^{+0.1}_{-1.6}$\& $13.4^{+2.0}_{-0.6}$      & $0.01\pm0.02$ \&$-0.07\pm0.04$ &     $0.46\pm0.05$ \&$0.38\pm0.06$        \\
NGC 4365  & -	           & 0.090	        & S &KDC    & -& -&  -     \\
NGC 4374  &$-0.40$ & 0.023  &S &SC      &$14.7^{+1.4}_{-1.3}$\& $16.1^{+1.6}_{-2.1}$       & $0.01\pm0.04$ \&$-0.15\pm0.04$&      $0.30\pm0.04$ \&$0.31\pm0.06$     \\
NGC 4382  &$0.59$   & 0.155  &F &CLV    &$3.7^{+0.7}_{-0.2}$\& $5.1^{+0.8}_{-0.2}$           & $0.051\pm0.04$ \&$-0.09\pm0.04$&     $0.16\pm0.04$ \&$0.20\pm0.04$        \\
NGC 4406  &  -            & 0.060    & S &KDC    & -  & - & -        \\
NGC 4473  &$1.03$   & 0.195  &F &MC      &$11.7^{+0.6}_{-1.0}$\& $12.2^{+1.8}_{-0.6}$      & $0.09\pm0.02$ \&$-0.09\pm0.04$&        $0.24\pm0.04$ \&$0.27\pm0.04$    \\ 
NGC 4486  &$-0.07$ & 0.019   & S&SC     &$17.7^{+0.1}_{-0.1}$\& $17.7^{+0.1}_{-0.1}$      &$0.13\pm0.02$ \&$-0.03\pm0.02$ &         $0.41\pm0.05$ \&$0.44\pm0.05$   \\
NGC 4526  &$-1.92$ & 0.476   &F &MC     &$6.4^{+0.3}_{-0.8}$\& $10.7^{+1.0}_{-0.9}$        & $0.19\pm0.02$ \&$-0.01\pm0.02$&         $0.22\pm0.04$ \&$0.28\pm0.05$   \\
NGC 4552  & 0.00      & 0.049   &S &KDC  &$10.2^{+1.0}_{-0.9}$\& $12.8^{+1.2}_{-1.1}$      & $0.21\pm0.04$ \&$0.03\pm0.04$&           $0.25\pm0.03$ \&$0.26\pm0.05$   \\

NGC 4570  &$1.90$  & 0.561   & F &MC   &$11.7^{+1.1}_{-1.0}$\& $14.1^{+1.4}_{-1.2}$       &$0.17\pm0.02$ \&$-0.13\pm0.04$ &          $0.17\pm0.03$ \&$0.27\pm0.04$  \\
NGC 4621  & 1.66      & 0.268   & F&KDC  &$13.4^{+1.3}_{-0.6}$\& $14.1^{+1.4}_{-1.2}$       & $0.09\pm0.04$ \&$-0.11\pm0.04$&          $0.28\pm0.05$ \&$0.32\pm0.06$   \\
NGC 4649  &-	           & 0.130           & F & -       & -&   -   &-   \\
NGC 4660  &$0.66$  & 0.472   &  F&MC    &$12.2^{+1.2}_{-0.6}$\& $13.4^{+1.3}_{-1.2}$       & $0.15\pm0.02$ \&$-0.11\pm0.02$&          $0.24\pm0.04$ \&$0.29\pm0.05$  \\
\hline
\end{tabular}
\caption{SAURON survey kinematic and stellar population parameters: (1) galaxy, (2) ``diskyness'' parameter ($a_4$), (3) rotation parameter ($\lambda_R$) and
(4) Slow or Fast rotator (\citealt{emsellem07}, \citealt{cap07}) according to (3), (5) kinematic group (GRP) (\citealt{emsellem07}: SC = single component, MC = Multiple
Component, KDC = Kinematically Decoulpled Core, CLV = Central low-level velocity), (6) Age estimates in Gyr within a circular aperture of Re/8 and Re respectively, (7) Metallicity estimates within a circular aperture of Re/8 and Re respectively and (8) $[\alpha/Fe]$ ratios. The last three columns are taken from \cite{kunt10} and are SSP-equivalent estimates.}
\label{sauron}
\end {scriptsize}
\end{table}
\end{landscape}

\begin{landscape}
\begin{table}
\begin{scriptsize}
\centering
\begin{tabular}{c c cccc  ccccccc}

\hline
(1)         & (2)            & (3)       &             (4)       &      (5)         &    (6)       &        (7) &          (8)        &            (9) &            (10) &           (11)            & (12)     & (13)  \\
\hline
Galaxy & date & total  &  effective  & seeing (\arcsec)& photometric &  aperture  &                    $A_K$       &     $A_g$   & $A_z$ &$R_{\rm{blend}}$ & $\rm N_{blend}$ &N\\
            &           &    exposure time (s)      &airmass    &                     & correction  & correction&                   &         & &($ACS_{\rm{pix}}$)&\\
\hline
NGC 3377& 18, 21 and 22 /03/2008 &12000  &1.334 &  $\sim$1.2 &$-$0.060$\pm0.007$   &  $ 0.286\pm0.012 $  & 0.013& 0.123      &  0.050  &30&10 &75  \\
NGC 4278& 22 and 24 /03/2008 &12000 & 1.586 &      $\sim$1.2 & -   &  $0.524\pm0.023 $  & 0.010 &  0.105    &    0.043   &30 &6&67\\
NGC 4365 & 27 and 28/02/2007 &2700 &   1.203&      $\sim$0.9  &-      &  $0.083\pm0.005$ & 0.008     & 0.076     &   0.0312   &25&12&99\\
NGC 4374 & 18 and 19 /03/2008&12000& 1.223 &       $\sim$1.5  &-      &  $ 0.348\pm0.015$  &0.015    &  0.145    &      0.059  &35&11&93\\
NGC 4382& 13 and 15 /03/2009 &12900 & 1.080 &      $\sim$1.0 &$-$0.18$\pm0.03$  &  $0.341\pm0.018 $  & 0.011  & 0.113     &     0.046  &25&5 &59\\
NGC 4406 & 28/02/2007 & 11700 & 1.206  &           $\sim$0.9  &-      &  $0.148\pm0.006$ & 0.011     &  0.109    &     0.044   &  25&8&76\\
NGC 4473&13, 14, and 15 /03/2009 &12600 & 1.436 &  $\sim$1.2 &$-$0.23$\pm0.02$   &  $0.351\pm0.016 $  & 0.010 &  0.102    &     0.041 &30&2&55 \\
NGC 4486 & 27 and 28/02/2007 &12000 &  1.195&     $\sim$1.1  &-      &  $0.190\pm0.008$ & 0.008      & 0.080     &    0.033   &30&76&301\\
NGC 4526 & 18 and 19 /03/2008& 12600&1.434 &       $\sim$1.4 &$-$0.050$\pm0.007$   &  $ 0.448\pm0.187$  &0.008  & 0.080     &    0.033 &35 &6 &55\\
NGC 4552 & 18 and 21 /03/2008 &12600 &1.253 &      $\sim$1.3  &$-$0.090$\pm0.006$   &  $ 0.233\pm0.009$  &0.015 & 0.149     &   0.061  & 30&6&107\\
NGC 4570& 15 and 16 /03/2009 &11400 & 1.242 &      $\sim$1.0&$-$0.20$\pm0.01$ &  $0.273\pm0.013 $  & 0.080    & 0.080     &    0.033   &  25&1&19\\
NGC 4621 &18 and 21 /03/2008 &12000 & 1.249&       $\sim$1.3 &$-$0.020$\pm0.001$  &  $ 0.270\pm0.011$  & 0.012 & 0.120     &   0.049  & 30 &10&80\\
NGC 4649 & 27 and 28/02/2007 &11700 & 1.284 &     $\sim$0.9  &-      &  $0.161\pm0.007$ & 0.010      &  0.094    &   0.039  &  25&21&164\\
NGC 4660& 15 and 16 /03/2009 &12300 & 1.258 &      $\sim$1.2  &- &  $0.466\pm0.020 $  & 0.012        & 0.120     &   0.049  &  30&6&51\\

\hline
\end{tabular}
\caption{Summary of the Observations - LIRIS K-band imaging and relevant calibration parameters for LIRIS and ACS: (1) galaxy, (2) dates the data were taken, (3) total exposure time in seconds, (4) effective airmass, (5) seeing, (6) value of photometric correction, (7) value of aperture correction, (8), (9), (10) galactic extinction in K($A_K$), g($A_g$) and z($A_z$) respectively, (11) blending radius ($R_{blend}$) (12) number of blended sources and (13) final number of clusters that satisfy the criteria summarised in Sect.\,5.}
\label{jobsliris}
\end {scriptsize}
\end{table}
\end{landscape}

\section{Observations and data reduction}

\subsection{LIRIS K-band imaging}
Imaging in the Ks-band (from now on referred to as K)  was obtained during 2007, 2008 and 2009 with the LIRIS near-infrared spectrograph and imager (\citealt{apulido03}, \citealt{manchado04}) on the William Herschel Telescope (WHT) in La Palma, Spain. 
A summary of the LIRIS observations is given in Table \ref{jobsliris}. This instrument has a field of view of 4.2$\arcmin$ X 4.2$\arcmin$ and a plate scale of $0.25\arcsec\,pix^{-1}$. We adopted a dithering pattern that consisted of taking exposures in 5 different positions across the detector, starting at the centre, moving to the upper right quadrant and in a clock wise rotation pattern spanning all the 4 quadrants with a jitter from 12\,$\arcsec$ to 20\,$\arcsec$. We took 4 exposures in each of the 5 positions of the detector. Each exposure consisted of 15 seconds to cope with the high count rate from the background in the K-band. The total exposure time reached for each galaxy is also given in Table \ref{jobsliris}.

Even though the fields are dominated by the host galaxy, we did not adopt the traditional method of sky subtraction in the NIR for crowded fields. In that method one takes a sky exposure for every science exposure and spends the same amount of time observing the sky as observing the actual target.
We only took a few sky exposures ($\sim25$ frames per galaxy), and these few frames allowed a fairly good sky subtraction on a subset of the sky exposures and a good enough ellipse modelling and removal of the galaxy light. The images resulting from this operation were then used to create the actual "sky" image.

For the 2007 run, 8 standard stars were observed during the night of 27/02 and 7 during the night of 28/02, and both nights were found to be photometric. For the 2008 run, 7 standard stars were observed during the night of 18/03, the only photometric night. For the 2009 run, 6 standard stars were observed during the night of 15/03, the only night considered to be photometric.
During the last run in 2009, we had problems with \textit{Calima} dust storms from the Sahara desert, which severely affected the data taken on the night of March 13. 

\subsection{LIRIS data reduction and calibration}
The data reduction was performed with the standard packages in IRAF and the LIRISDR package written by Jos\'e Acosta Pulido and kindly provided to us by the author. The shifts between exposures were determined by measuring the centroids of 2 (min) to 4 (when possible) isolated bright stars in each exposure. 
The first step, was the correction of the pixel-mapping anomaly of LIRIS with LCPIXMAP in LIRISDR before any reduction procedure took place.
When the two-dimensional image is reconstructed from a one dimensional array, some of the pixels in the LIRIS image end up in places where they should not be. This affects mainly the lower left of the first quadrant (\citealt{cookbook}). 

A crude sky subtraction using sky frames within $\lesssim~30$\,min from the science exposures was performed. The resulting images from this crude sky subtraction were then combined after flat-field correction in each individual exposure. From this combined image an ellipse model was created with the STSDAS tasks ELLIPSE and BMODEL. This ellipse model was then multiplied by the flat field, shifted and subtracted from each original exposure. We were left with exposures without the central galaxy cusp on them. These "galaxy-free" images were median combined following the standard procedure of sky subtraction in the near-infrared: a "running mean" of the 4 exposures obtained closest in time (2 before and 2 after) was constructed for each frame. The result of this operation yielded the sky image to be subtracted from each original exposure. This technique of sky subtraction has optimal results for the simplest E0 galaxies, but as the galaxy morphology becomes more complex, the ellipse model fit becomes poorer. Fairly good ellipse models are obtained for all types of E-type galaxies but poor models are achieved for S0 galaxies as these galaxies contain an additional disc component. Since ELLIPSE is not able to model such a component, in the disc-dominated part of the galaxy our photometry will be compromised.  

\begin{figure*}
\resizebox{\hsize}{!}{\includegraphics[]{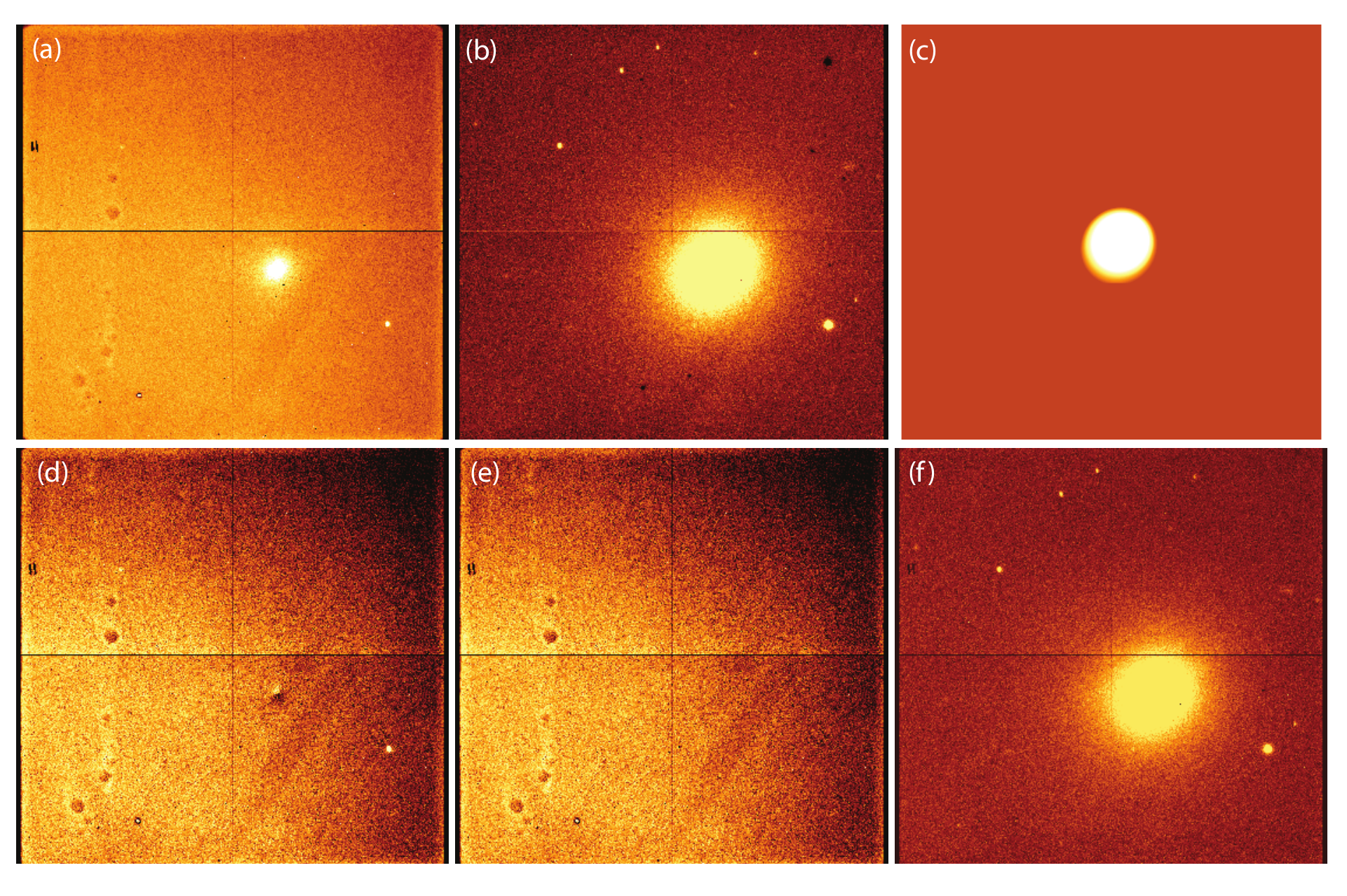}}
\caption[]{Ilustration of some of the important steps involved in the sky subtraction for NGC\,4374.
(a): science frame, (b): rough sky subtracted science frame, (c): ellipse model, (d): ellipse model subtracted frame, (e): an average sky frame, (f): sky subtracted science frame}
\label{modsubilust}
\end{figure*}

In Fig. \ref{modsubilust} we illustrate some of the steps of the sky subtraction. Panel (a) shows a 15-second science frame that is sky subtracted by the few science frames (taken within $\lesssim30$\,min from it) and results in panel (b). A stack of images like panel (b) are averaged together ($60-200$ frames), and the resulting image is used to create an ellipse model fit, shown in panel (c). If one subtracts an image like (c) from an image like (a), the result is (d). Median combining several images like (d) with different dither positions will allow for the removal of its model residual, seen in the centre of this panel as well as any bright foreground star or background galaxy that is in the image. The result of this combination will be an image like the one shown in panel (e), an average sky frame. Subtracting the constructed sky from the original science frame (a) will result in a sky subtracted science frame, shown in panel (f). 

\begin{figure*}[ht]
\resizebox{\hsize}{!}{\includegraphics[]{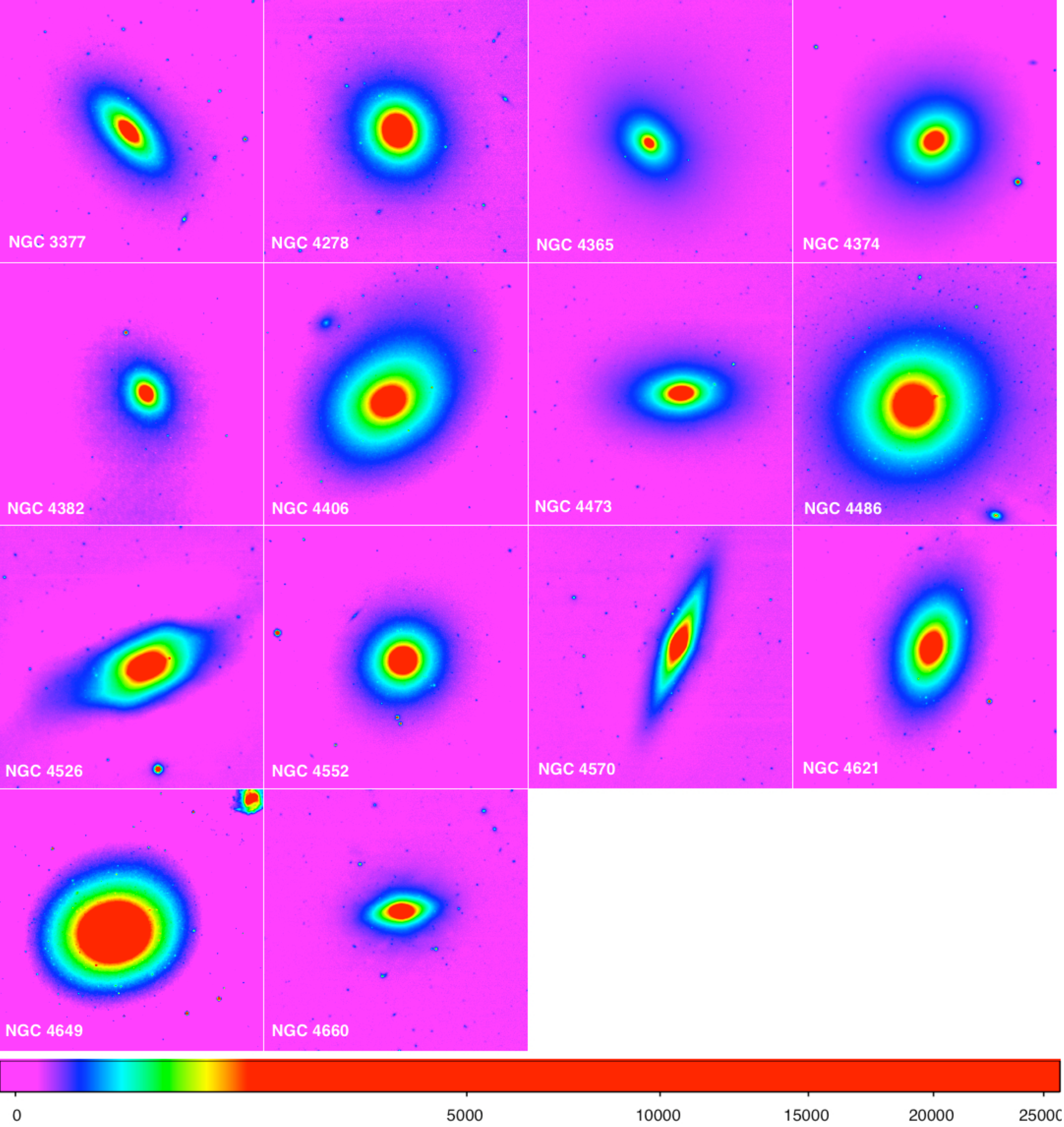}}
\caption[]{Liris final reduced K-band images for the 14 galaxies. North is up and east to the left. Each image is $\sim\,270\,\arcsec$ on a side, corresponding to $\sim\,24$\,Kpc for Virgo Cluster galaxies, $\sim\,21$\,Kpc for NGC\,4278 and $\sim\,15$\,Kpc for NGC\,3377. The colour bar is given in units of counts.}
\label{images}
\end{figure*}

After flat-fielding the sky-subtracted images we corrected for the vertical gradient with LICVGRAD in LIRISDR, another anomaly suffered by LIRIS. Some exposures present a discontinuous jump between the upper and the lower two read-out quadrants, with the amplitude of this jump depending on the order of an image in a series of continuous exposures  (\citealt{cookbook}). This happens because when the telescope is moved from one dither position to another it takes about 3 exposures for the detector to reach equilibrium. Each image is then corrected individually.
Following the vertical gradient correction, the images were shifted and averaged resulting in a reduced, final image. Figure \ref{images} shows these images for our galaxy sample. Another elliptical model was created for each galaxy and subtracted from the final, reduced image. We did not use the same model constructed through a crude sky subtraction (eg. Fig. \ref{modsubilust} panel (c)) as it is based on a small fraction of the images, and the final, reduced images contain $\sim\,800$ exposures. In this model-subtracted image we performed photometry with PHOT after object detection in the optical ACS images, as described in Sect. 3.3.

To calibrate our data photometrically, we assumed a transformation of the following form:
\begin{equation}
K_s= k_s + z_k - p_k \times x_{k_{\rm{eff}}}
\end{equation}
where $K_s$ and $k_s$ are the standard and instrumental magnitudes, and $p_k$ and $x_{k_{\rm{eff}}}$ are the atmospheric extinction coefficient and the effective airmass, respectively. 
To determine $p_k$ we needed a broad range in airmass values, and the standard star exposures didn't span the needed range. We then used both standard stars and individual galaxy exposures that contained bright stars taken at different times during the night, ie., covering sufficient airmass values to derive $p_k$ during the 2007 run. An estimate of $p_k$ was calculated to be the slope of the linear relation between airmass and the observed magnitudes of the same standard star and/or bright stars in the individual science frame exposures taken during the night. We adopted a value of $p_k$ = 0.07 $\pm$ 0.02$\,\rm(unit~airmass)^{-1}$, determined as the mean of the slopes found for different bright and/or standard stars in these exposures. There were no previously documented values for $p_k$ at this wavelength for La Palma. Typical values found in the literature for $p_k$ for sites such as Cerro Paranal and La Silla are $p_k$ = 0.05 $\rm(unit~airmass)^{-1}$ (\citealt{puzia02}, \citealt{lbs05}) and  $p_k$ = 0.08 $\rm(unit~airmass)^{-1}$ for Las Campanas (\citealt{hempel07}).
To determine $x_{k_{\rm{eff}}}$, the mean values for combined exposures, we used the following expression, which accounts for the airmass in each of the individual exposures and where the subscript $i$ designates the $i^{th}$ exposure
\begin{equation}
x_{\rm{eff}}=-\frac{log[\frac{1}{N}\sum_{i=0}^N 10^{-0.4(p_k \times x_{k_{i}})}]} {0.4 \times p_k}.
\end{equation}
Equation (2) is found by summing up the individual contributions of each airmass value in the equality, and solving for $x_{\rm{eff}}$
\begin{equation}
F_{\rm{obs}}=F_{k} \times 10^{-0.4 p_{k} \times x_{\rm{eff}}}.
\end{equation}
Following the standard star observations, we determined the zero point using an aperture of 20 pixels. For the 2007 run it was found to be $z_k=23.450 \pm 0.004$, for the 2008 run  $z_k=23.110 \pm 0.004$ and for the 2009 run $z_k=23.070 \pm 0.008$.  For the 2008 and 2009 runs only the first night was found to be photometric and an extra step, ie. the photometric correction, was added to the equation. These values are listed in Table \ref{jobsliris}. The errors in the $z_k$ values are the standard errors of the mean.
Objects were detected in the ACS images (see Sect. 3.3) and transformed to the LIRIS coordinate system. Aperture photometry was performed with PHOT in a 5 pixel aperture in the final, model-subtracted image.
We estimated photometric corrections by measuring the magnitudes of some bright objects ($\sim10$) for each galaxy in a combined image with all exposures and only with the photometric night exposures. The weighted mean difference between these for each galaxy was taken to be the photometric correction adopted value.
An effort was made to always obtain some exposures of each galaxy under photometric conditions. The only galaxy where this was not possible was NGC\,4278 with exposures taken on the nights of March 22/03/08 and 24/03/08. An analysis of the relation between airmass and instrumental magnitude for a few bright objetcs in all exposures showed that even though the latter night was not photometric, its exposures were not as severely affect by extinction as for the first night. An estimate for the extinction, through exposures taken in the first night, was determined to be $0.030\pm0.006$\,mag. Since this value is small and we are not certain about the extinction for the last night, we did not apply the photometric correction for this galaxy.  
As can be seen in Table \ref{jobsliris}, the uncertainties in the photometric corrections are $\le 0.007$ for the second run and $\le 0.03$ for the third one.
Aperture corrections were calculated with APPFILE from 5 to 20 pixels and cross-checked by determining the difference between the magnitudes measured with a 5 pixel aperture to a 20 pixel one, and the values proved to be consistent.
With the uncertainties on the zero points, photometric and aperture corrections listed above, we estimate the calibration to be accurate to $\sim 0.04$ for M\,87, $\sim 0.03$ for NGC\,3377 and  $\sim 0.05$ for NGC\,4660. We show these estimates as examples for each of the observation runs. These accuracy estimates were calculated by adding in quadrature the uncertainties in the zero points, photometric and aperture corrections. 
The magnitudes were corrected for galactic extinction using the values in NED\footnote{This research has made use of the NASA/IPAC Extragalactic Database (NED) which is operated by the Jet Propulsion Laboratory, California Institute of Technology, under contract with the National Aeronautics and Space Administration.} from \cite{schlegel98}. These quantities are also listed in Table \ref{jobsliris}.

\subsection{HST ACS imaging}
ACS pipeline processed images of the 14 galaxies were downloaded from the MAST archive. The data for 12 of them were obtained as part of the ACS Virgo Cluster Survey (ACSVCS; \citealt{cote04}) and use the F475W ($\sim$Sloan g) and  F850LP ($\sim$Sloan z) filters. 
The exposure times were 750\,s and 1210\,s in F475W and 
F850LP, respectively, split into 2 and 3 exposures for cosmic-ray rejection. From now on we refer to these filters as g and z.  
For the galaxies not located in the Virgo Galaxy Cluster, NGC\,3377 and NGC\,4278, there were 1 and 2 pointings, respectively, that covered the LIRIS field of view. The ACS Program IDs are 10435 and 10835. For the former the exposure times were 1380\,s for g and 3005\,s for z, divided into 4 exposures for both g and z. For the latter galaxy, the exposure time for the z band was 1200s for the two pointings, split into 3 exposures; for the g band the central pointing has 676.2\,s of exposure time and for a west pointing 702\,s split into 2 exposures. 

The data were obtained with the wide field channel on ACS, which has a pixel scale of $0.050\arcsec\,pix^{-1}$ and a field of view of $3.36\arcmin \times 3.36\arcmin$. 
This is roughly the field of view of the final, combined LIRIS images considering the dithering pattern. 
This gives us a radial coverage of $\sim8\,kpc$ for NGC\,3377, $\sim11\,kpc$ for NGC\,4278 and $\sim12\,kpc$ for the Virgo Cluster galaxies.
In terms of the effective radius ($R_{e}$) of the galaxies, it is  $0.29R_{e}$ for NGC\,3377, $0.22\,R_{e}$ for NGC\,4278 and for the Virgo Cluster galaxies $0.08\,R_{e}$ for NGC\,4486 and $1.3\,R_{e}$ for NGC\,4660.
They were then combined with the MULTIDRIZZLE task in the STSDAS.DRIZZLE package, which removes the geometric distortion and normalises the images to exposure times of 1\,s. A $20 \times 20$ median filter image was subtracted from the multidrizzled image and the sources were detected on the resulting image with DAOPHOT in the z band image. 
Since there is significantly more background noise near the centre of galaxies than farther out, more spurious objects are detected in the central part of the galaxy. At this point nothing was done to remove these spurious objects as they are expected to be automatically removed from the sample with the application of the criteria to refine the GC selection, explained in Sect.\,4.
Aperture photometry was performed in a 5-pixel aperture. The following aperture corrections were used for the Virgo Cluster galaxies: $-$0.09 in g and $-$0.15 in z from 5 to 10 pixels. These values were taken from Strader et al. (2006), who derived them as median corrections from bright objects in the five most luminous galaxies of the Virgo Cluster Survey. 
For NGC\,3377 and NGC\,4278 the aperture corrections from a 5 to 10 pixel radius were determined with APFILE and found to be $-$0.132 in g and $-$0.170 in z for the former and $-$0.09 in g and $-$0.13 in z for the latter. For all 14 galaxies the final correction from 10 pixels to infinity was $-$0.10 in g and $-$0.12 in z, from Table 5 of  \cite{sirianni05}.   
Extinction corrections were applied according to $A_{g}=3.634 \times E(B-V)$ and $A_{z}=1.485 \times E(B-V)$, where the reddening $E(B-V)$ value is from NED and the ratios correspond to the spectral energy distribution of a G2 star from Table 14 of \cite{sirianni05} and are appropriate for GCs (\citealt{jordan04}). The final values $A_g$ and $A_z$ for each galaxy are listed in Table \ref{jobsliris}. 
The magnitudes were finally transformed to the AB system using as zero points for g and z, 26.068\,mag and 24.862\,mag from Table 10 of \cite{sirianni05}.

\subsection{Comparison with ACSVCS}
\begin{figure}
\resizebox{\hsize}{!}{\includegraphics[angle=90]{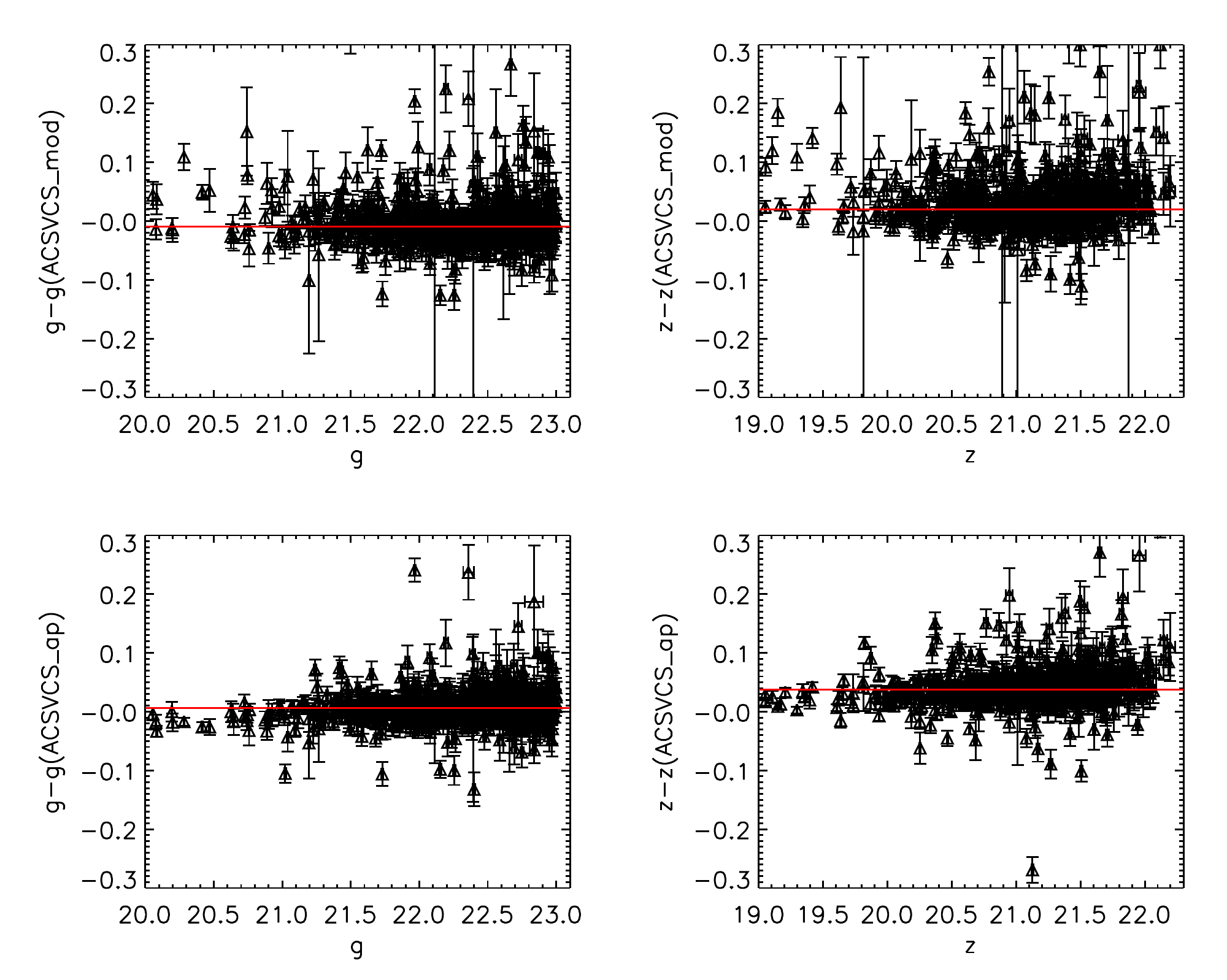}}
\caption[]{Comparison of the ACS magnitudes of the present study and those from the ACSVCS for the combined sample of NGC\,4486 and NGC\,4649. The solid red lines indicate the median differences.
The top panels are for the \textit{model magnitudes} and the bottom panels for the \textit{average correction aperture magnitudes} of the ACSVCS.
}
\label{acsvcs}
\end{figure}

We have not used the published catalogues of GCs of the ACSVCS (\citealt{jordan09}) as they were not available when the present study began.
Nevertheless, we compare the $g$ and $z$ magnitudes of this study with those of the ACSVCS for the combined sample of NGC\,4486 and NGC\,4649.
The $\sim500$ sources of these two galaxies before visual inspection were matched with the ACSVCS catalogue.
In Figure \ref{acsvcs} the difference between the $g$ and $z$ magnitudes of the present study and those of the ACSVCS are shown as function of the $g$ and $z$ magnitudes of the present study.
The top panels are for the \textit{model magnitudes} and the bottom panels for the \textit{average correction aperture magnitudes} of the ACSVCS (see \citealt{jordan09}).
The median difference between the $g$ magnitudes is $\sim-0.009$ for the model and $\sim0.006$ for the average correction aperture magnitudes.
On the other hand, the difference between the $z$ magnitudes is $\sim0.030$ for both comparisons, slightly larger for the second case.
The scatter is smaller for the comparisson with the average correction aperture magnitudes of the ACSVCS, as expected, once the same method of average correction obtained from bright GC candidates is adopted in both cases.
The ACSVCS model magnitudes include the correction for the size of each cluster.
We conclude that either we have underestimated the aperture correction for the z band or that the ACSVCS study has overestimated it. 
This difference however, is not expected to affect significantly the results of this series of papers. 

\subsection{NGC\,4365: LIRIS vs. SOFI comparisson}
\begin{figure}
\resizebox{\hsize}{!}{\includegraphics{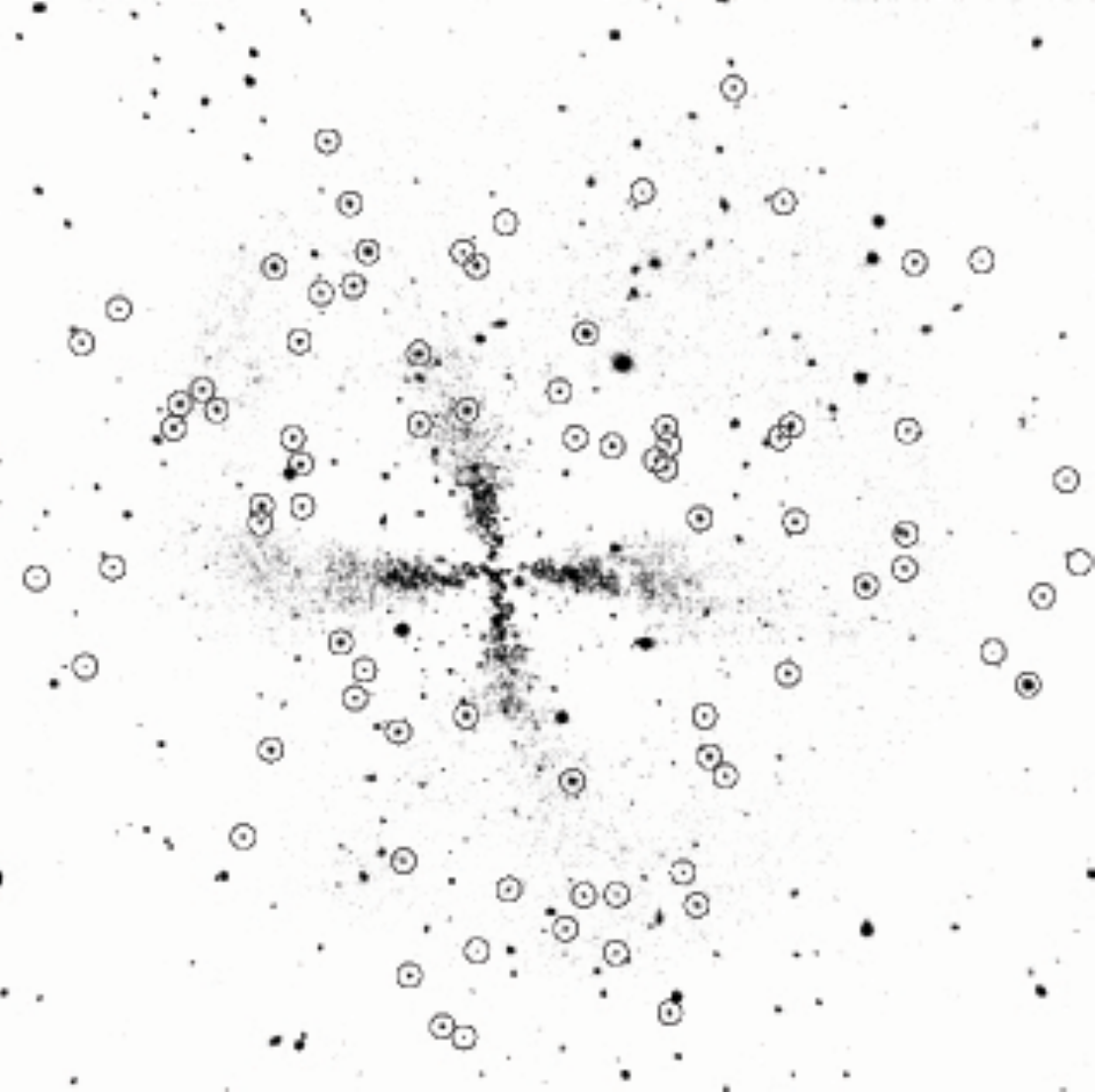}}
\caption[]{WHT/LIRIS K-band image of NGC\,4365. The 78 objects in common with NTT/SPFI from \cite{lbs05} are marked. North is up and east is left; the grey scale is inverted. The cross-shaped feature  is a residual from the galaxy model subtraction.}
\label{n4365image}
\end{figure}

\begin{figure}
\resizebox{\hsize}{!}{\includegraphics[angle=90]{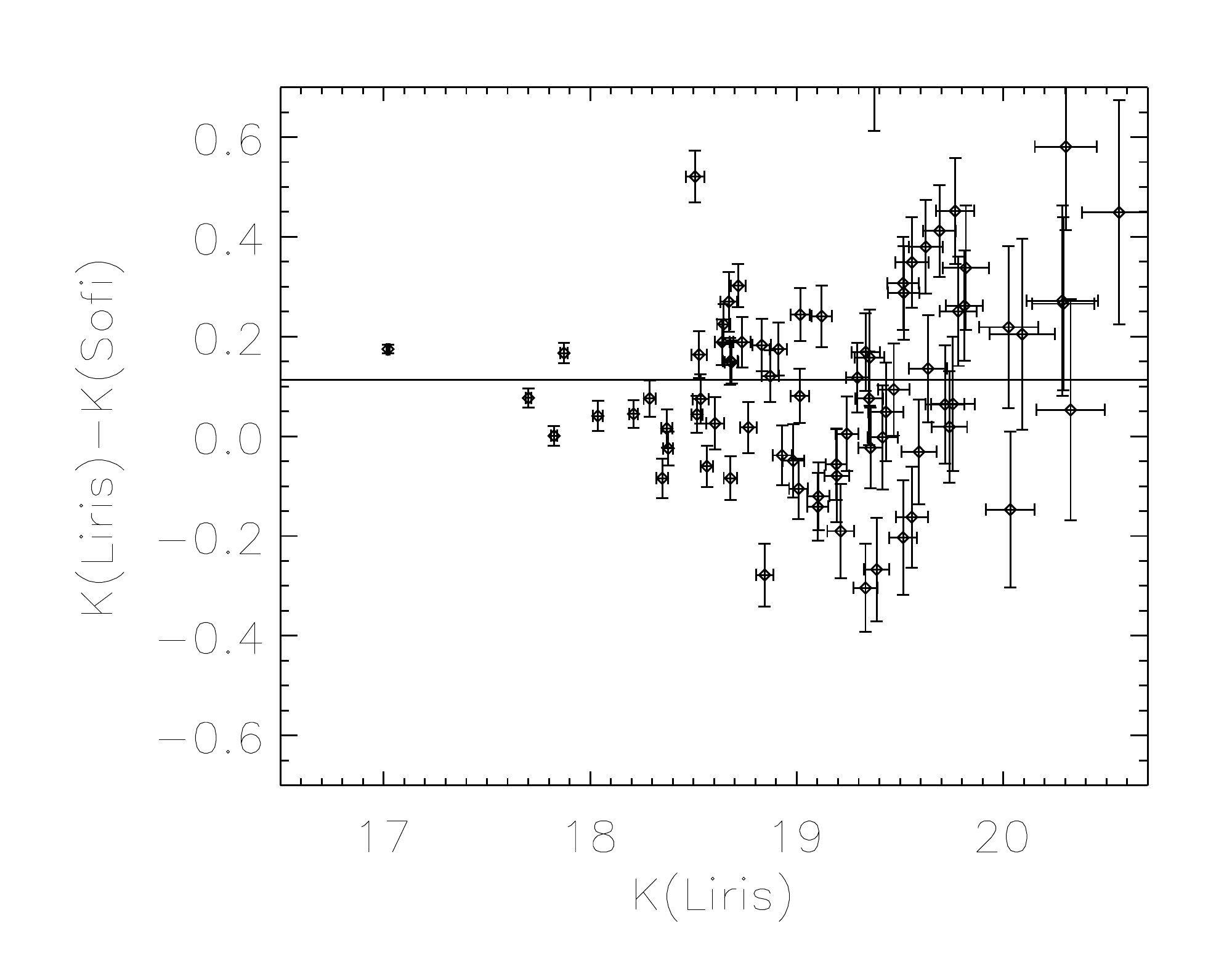}}
\caption[]{Comparison of WHT/LIRIS and NTT/SOFI GC magnitudes. The weighted mean difference is $0.113\pm0.005$, and the solid line indicates this value. There were 78 matches.}
\label{sofimatchliris}
\end{figure}

LIRIS imaging was taken for the galaxy NGC\,4365 which showed evidence for having intermediate age clusters in previous optical/NIR photometry (with ISAAC/VLT, \citealt{puzia02}) and was thoroughly studied by \citealt{lbs05}. We compared the NTT/SOFI GCs for NGC\,4365 from \cite{lbs05} and found 78 objects in common, which are shown in Figure \ref{n4365image}. Figure \ref{sofimatchliris} shows a comparison of the LIRIS and SOFI K-band photometry, with the weighted mean difference being $0.113\pm0.005$ for all matched sources.
It seems that the difference between the LIRIS and SOFI photometry is magnitude dependent as the differences become larger as on goes to fainter magnitudes, with the exception of the brightest magnitudes ($\sim$17, 18 mag). It is difficult to understand the origin of this trend which is opposite to what is seen for the SOFI and ISSAC comparison (\citealt{lbs05}) where magnitude differences become smaller with fainter magnitudes. If one removes the object with K(LIRIS)\,$\sim$\,17, the brightest of the 5 objects located close to the right border of Fig. \ref{n4365image}, the trend does not appear as significant.
The mean and median differences between LIRIS and SOFI photometry are $0.139\pm0.031$ and $0.094$ respectively. When considering only sources brighter than 20 mag, the mean and median values are $0.09\pm0.02$ and $0.075$, respectively, while the weighted mean is $0.112\pm0.005$.
The difference found by \cite{lbs05} between the ISAAC (\citealt{puzia02}) and SOFI data was $0.102\pm0.026$.
Thus, the LIRIS magnitudes are similar to those from ISAAC, while the SOFI measurements show an offset of $\sim 0.1$\,mag.
In terms of ages and metallicities, this magnitude difference of $\sim 0.1$\,mag in the $K$-band means that the LIRIS and the ISAAC photometric systems will yield more metal-rich and younger GCs relatively to the SOFI system.
This offset, although large, is reasonable given the difficulties of accurately calibrating NIR photometry due to the bright sky background, uncertain photometric zero-points, and aperture and photometric corrections. For a more detailed discussion see \cite{lbs05}.

\section{K-band photometric accuracy}
The errors returned by PHOT might not be sufficiently accurate. They are determined from an image that had the galaxy light removed from, and when its individual exposures were combined, there was subpixel shifting. The combination of these two effects can be responsible for smoothing out the original noise fluctuations. When the photometric errors are estimated they will not account for these noise fluctuations or the diffuse galaxy light over the clusters. Therefore, the output errors could be underestimated. To estimate the \textit{true} uncertainty in the photometry, a couple of tests were performed. 
\subsection{Artificial star tests}
\begin{figure}[htbp]
\begin{center}
\resizebox{\hsize}{!}{\includegraphics[angle=90]{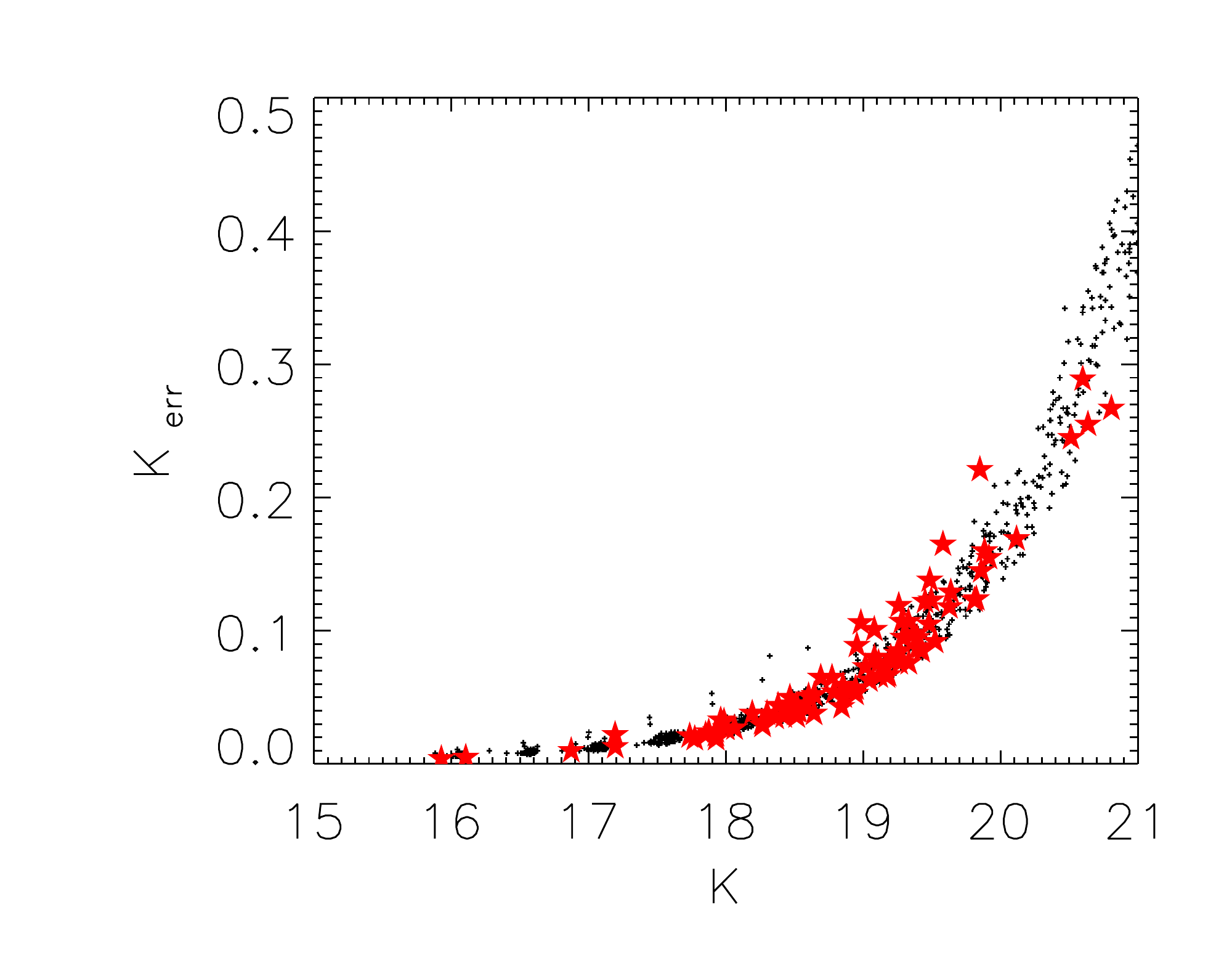}}
\caption{$K_{err}$ as measured by PHOT as a function of  $K$ for GCs of NGC\,4278 marked as stars and for the artificial star test output objects marked with filled dots.}
\label{kerr_k}
\end{center}
\end{figure}

\begin{figure}[htbp]
\begin{center}
\resizebox{\hsize}{!}{\includegraphics{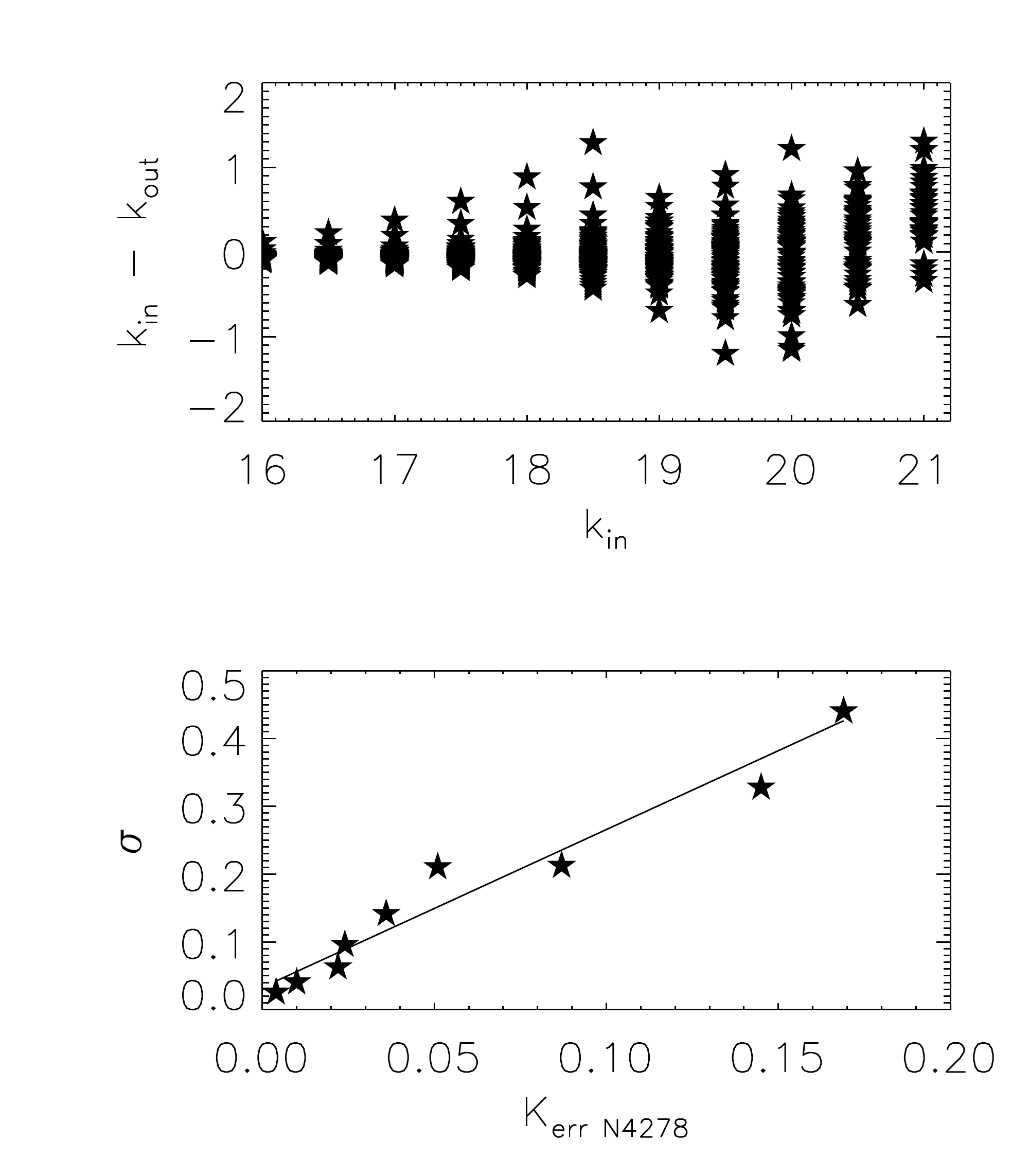}}
\caption{\textit{Upper panel:} input K binned in $0.5$ magnitude vs. the difference between the input K magnitude and the output magnitude. \textit{Bottom panel:} the output binned K PHOT error estimate for NGC\,4278 ($Kerr_{N4278}$) \textit{vs.} the estimated sigma for the corresponding binned magnitude. The best fit between these two quantities is plotted as a solid line. }
\label{startest}
\end{center}
\end{figure}

Artificial star tests with BAOLAB.MKSYNTH \citep{larsen99} were performed. At these distances, GCs are point like sources when observed from the ground. 
This task generates a synthetic image consisting of point sources with a shape based on the PSF of the image. The PSF was created with the IRAF PSF and SEEPSF tasks from bright stars in the NGC\,4278 K band image.
A grid of evenly spaced sources with the same magnitude was generated for $16<K<22$, binned in 0.5 magnitudes, and added to real images of one of the galaxies. 
This corresponds to a total of 110 sources added per magnitude interval, which adds up to 1430 sources. The grid was completely regular, and the separation between the added sources was 80 pixels. 
The galaxy image used for this was NGC\,4278 because its field is not particularly crowded and there were sufficient star-like objects to create a decent PSF. The seeing of this galaxy is $\sim1.2\arcsec$, while some other galaxies have better seeing (eg.  $\sim0.9$ for NGC\,4365, NGC\,4406 and NGC\,4649). With worse seeing the S/N drops, and therefore that the final photometric scatter that we derive will be overestimated for the cases with slightly better seeing.
Having images with point-like sources, we recovered the objects when possible and tested the difference between the input and the recovered magnitudes. The photometry was performed in the same way as for the real GC candidates in the galaxies.
We allow a maximum shift of 1 pixel from the input position when recovering the artificial objects and apply a cut in the error at $K_{\rm err} = 0.5$.
The last two cuts mean that we do not keep objects whose positions shift more than 1 pixel from the position where the artificial object was added. 
In Figure \ref{kerr_k} the relation $K_{\rm err}$, as measured by PHOT, as a funcion of $K$ is plotted as filled stars for the final GC sample of NGC\,4278, and the output $K_{\rm err}$ as a function of the K output magnitude for the artificial star test objects is plotted with filled points for the criteria: $K_{\rm err} < 0.5$, $x_{\rm sh} \leq -1, 1$ and $y_{\rm sh} \leq -1, 1$.
By comparing the GCs of NGC\,4278 in Fig. \ref{kerr_k} with the artificial star objects, 
it is readily visible that the relation between $K$ \textit{vs.} $K_{\rm err}$ for these two sets of objects resemble each other. Thus, applying these 2 criteria to the artificial star objects, we are able to reproduce fairly well the $K$ and $K_{\rm err}$ curve for NGC\,4278.
In the upper panel of Figure \ref{startest} the input K binned in $0.5$ magnitude vs. the difference between the input K magnitude and the output magnitude is shown. The magnitude difference increases as we go to fainter magnitudes, and we notice a positive offset in the difference around $K=20.5$. This offset is attributed to recovering a brighter object than it actually is. This can happen because when there is a brighter object nearby than the input artificial object, PHOT will tend to measure this brighter object, even when the maximum shift allowed is 1 pixel. By applying the $x_{\rm sh} \leq -1, 1$, $y_{\rm sh} \leq -1, 1$ criteria, these objects with larger position offsets were removed and thus the $K_{\rm in}\,-\,K_{\rm out}$ went from the maximum value of 5.2 to 1.4 for $K\,=\,21$. Another reason for this to happen is that the random fluctuations in the background are so high that, for faint objects, a positive fluctuation will make the object's estimated magnitude brighter than the input magnitude. 
Also, large negative fluctuations would result in a negative flux within the aperture and thus no magnitude would be measured. This is a reason why some objects are not detected. A small, negative fluctuation within the aperture could also result in a underestimation of the magnitude with $K_{\rm in}\,-\,K_{\rm out}\,<\,0$.  
In the bottom panel, the output binned K PHOT error estimate for NGC\,4278 ($K{\rm err_{N4278}}$) \textit{vs.} the estimated sigma for the corresponding binned magnitude is shown. The best-fit line between these two quantities is plotted as a solid line, and the relation is given by:

\begin{equation}
\sigma= 2.33\,\times\,K_{err_{N4278}} +0.03
\end{equation}

\subsection{SExtractor uncertainty determination}

The accuracy of the K-band photometry was also tested with SExtractor (\citealt{bertin96}) for one case. Unlike PHOT, SExtractor (\citealt{bertin96}) takes into account that there was a galaxy underlying the detected sources when determining the photometry, which means that another source of uncertainty will be added to the final photometric error budget. 
While the photometric uncertainties estimated by PHOT are based on the background annuli, SExtractor allows the use of two images: one with galaxy subtraction and another without it. This latter image is used to create a weighting image that is then used to calculate the background and thus the errors. As a test case we chose NGC\,4486. This galaxy has $\sim$\,300 clusters in the LIRIS/ACS field of view, and this will give a more robust comparison in terms of number of clusters. 
We ran SExtractor for NGC\,4486 and the program detected 77\% of the 301 sources that fullfill the criteria described in Sect. 5.

\begin{figure}[htbp]
\begin{center}
\resizebox{\hsize}{!}{\includegraphics{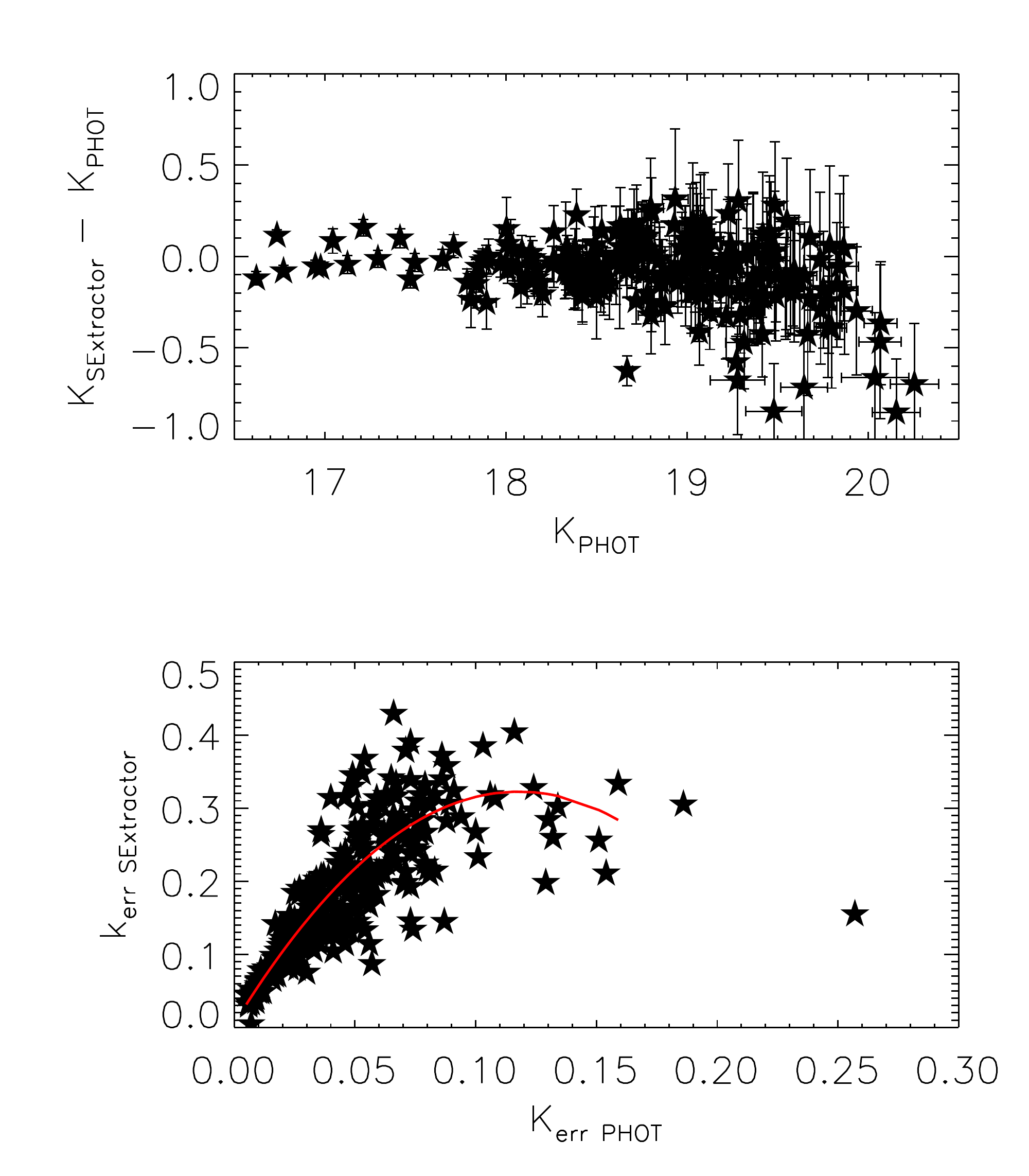}}
\caption{\textit{Upper panel:} K magnitude determined by PHOT for NGC\,4486 \textit{vs.} the difference between the magnitudes determined by SExtractor and PHOT. \textit{Bottom panel:} The magnitude error determined by PHOT for NGC\,4486 \textit{vs.} the magnitude error determined by SExtractor. The best second order polynomial fit between these two quantities is plotted as a solid red line.}
\label{photmatchsex}
\end{center}
\end{figure}

The upper panel of Fig.\,\ref{photmatchsex} shows the K magnitude determined by PHOT for NGC\,4486 \textit{vs.} the difference between the magnitudes determined by SExtractor and PHOT, and the weighted mean difference is $0.047\pm0.008$. Note that around $K_{\rm PHOT}\,\ga\,19$ there is an excess of sources with a $K_{\rm SExtractor} - K_{\rm PHOT}\sim-0.7$. These objects have particularly high negative background fluctuations within the annuli used to remove the sky contribution, which could be the cause of this negative feature. The bottom panel shows the magnitude error determined by PHOT for NGC\,4486 \textit{vs.} the magnitude error determined by SExtractor and the best second order polynomial fit between these, given by (5), is

\begin{equation}
K_{\rm err\,SExtractor}= -22.88\,\times\,K_{\rm err\,PHOT} +5.40\,\times\,K_{\rm err\,PHOT}^{2} 
\end{equation}

Note that the correlations between the uncertainties in the photometry determined by PHOT and the ones determined from artificial star tests and by SExtractor, given by Eqs. (3) and (4), respectively, are very similar.
The conclusion from these tests with BAOLAB.MKSYNTH and SExtractor is that PHOT errors are underestimated by about a factor of 2, and this has to be taken into account in any further analysis.

\section{Selection of globular cluster candidates}
The coordinates of the objects in the ACS $z$ frame were transformed to the LIRIS frame with the GEOMAP and GEOXYTRAN tasks in IRAF. We used as many objects as possible to compute the transformation, from a minimum of 10 to a maximum of 25. With GEOMAP the transformation from the ACS to the LIRIS coordinate system was computed with an output RMS of the transformation for all galaxies smaller than 0.25 LIRIS pixels. The ACS $z$-band coordinates were then transformed to LIRIS coordinates and aperture photometry was measured with PHOT with a centroid algorithm allowing a maximum shift of 1 LIRIS pixel. 
The differences between the coordinate systems of the $g$ and $z$ images were a small fraction of a pixel ($\leq 0.1$ ACS pixel). Running photometry in the $g$ image with the coordinates from the $z$ image was accurate enough.

A typical GC has a size of $\sim3$\,pc, and at the distance of the galaxies of the sample 1\,pixel\,$\sim$\,4\,pc. The spatial resolution of the ACS images allows classification of GCs by their spatial extent (they are somewhat more extended than stars at the ACS resolution), and it is thus straight-forward to eliminate most of the potential contamination from stars and/or galaxies. Effective radii ($R_{\rm{eff}}$) were measured in g images using the ISHAPE code described in \cite{larsen99}. The sizes were measured in these images due to the fact that the PSF in this band is more centrally concentrated. King models with concentration parameter $c$ ($c=r_{tidal}/r_{core}$), fixed to 30 were convolved with an empirical PSF derived from a 47Tuc g band ACS image taken around the same time the ACSVCS started. The PSF was created with stars in this 47Tuc image with the PSF and SEEPSF tasks on IRAF. To convert the measured sizes from ISHAPE into parsecs, galaxy distances from \cite{tonry01} were used for all galaxies with the exception of NGC\,4570, whose distance was chosen to be the mean of the available methods presented in NED (\citealt{jordan05} and \citealt{tully88}). All these distances are listed in Table \ref{sample}.
Only objects with a photometric measurement in all three bands ($g$, $z$ and $K$) were considered to be in the sample. 

\begin{figure*}
\includegraphics[width=3.5in]{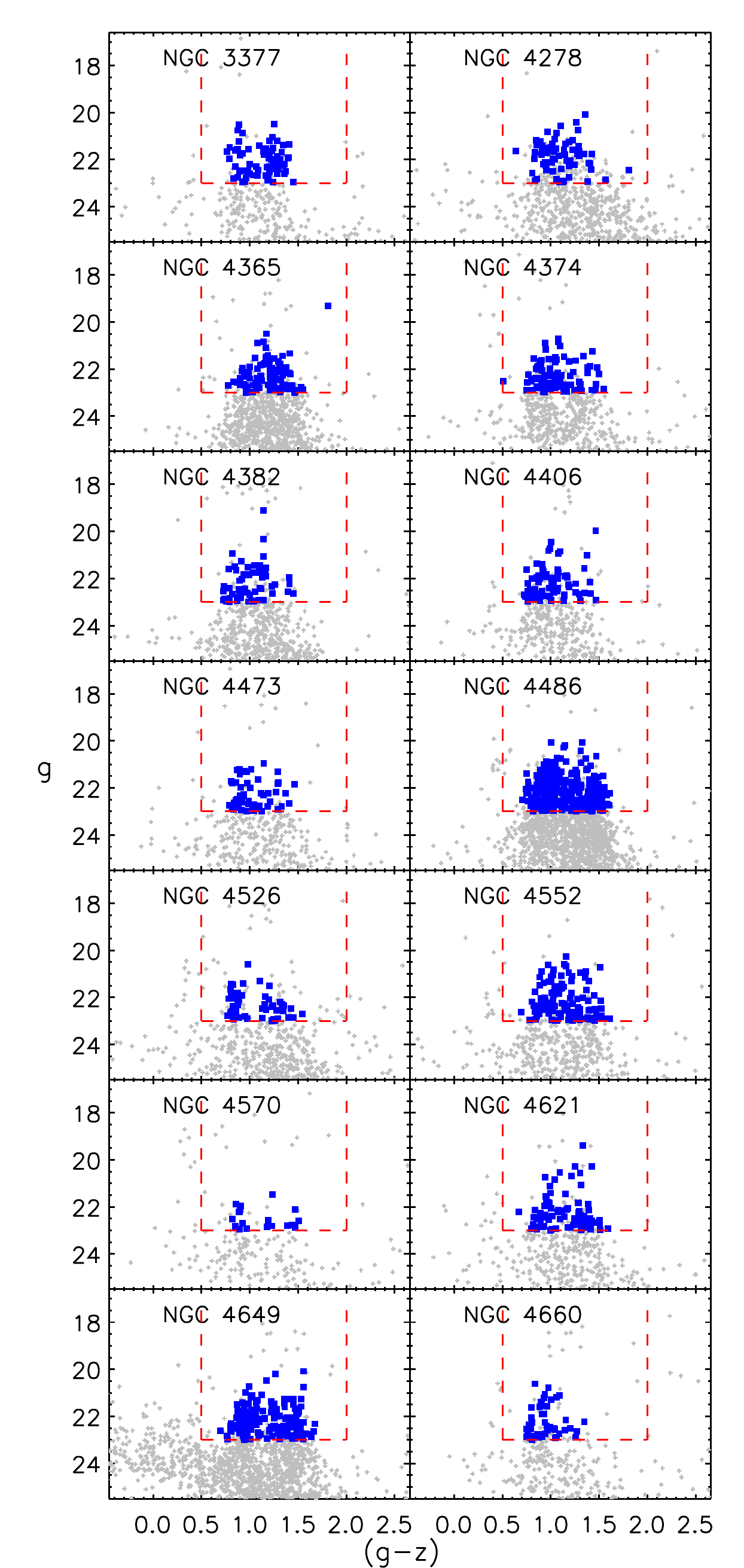}
\includegraphics[width=3.5in]{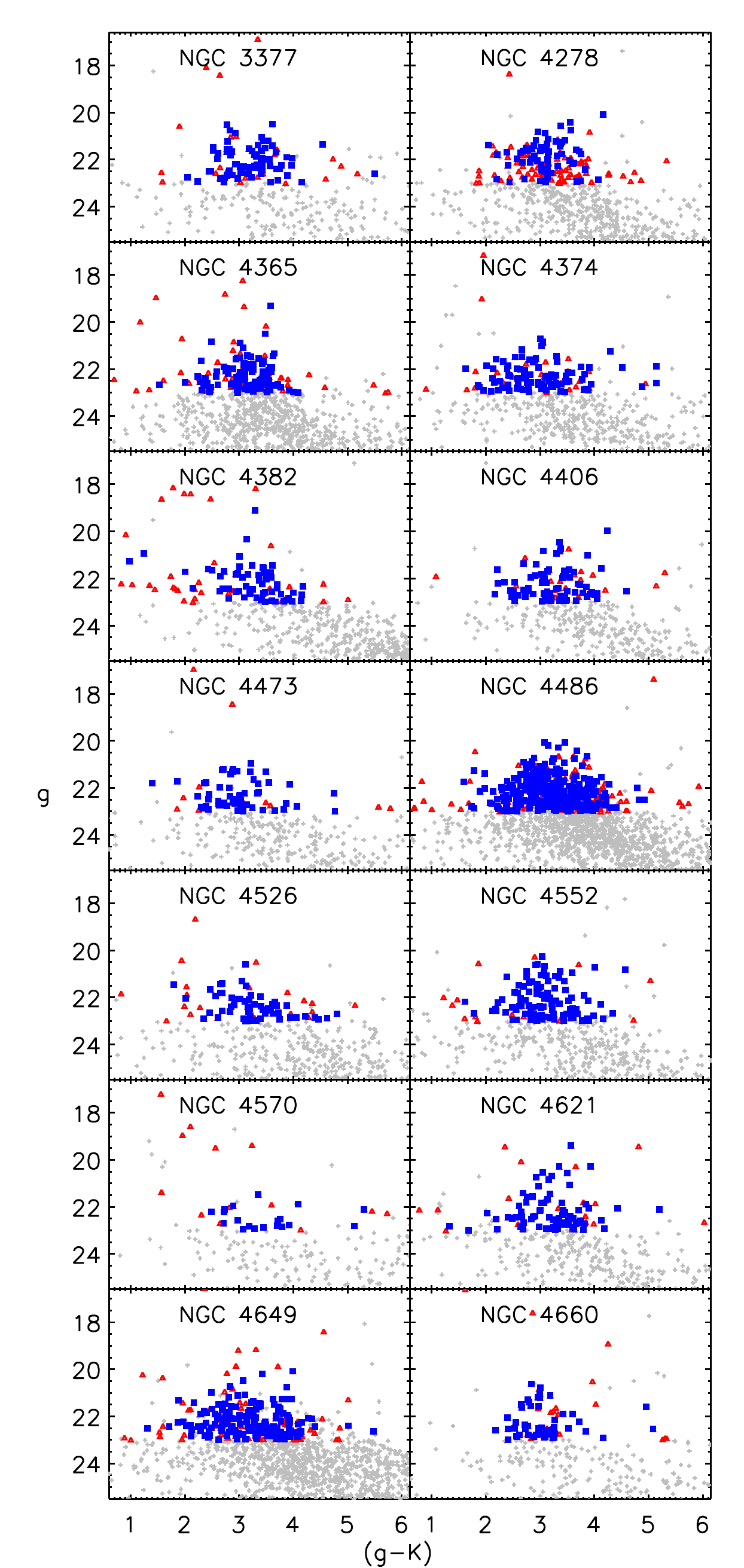}
\caption[]{\textit{Left panel: }Optical ($(g-z)\,vs.\,g$) colour magnitude diagrams for globular cluster candidates detected in g, z and K that had a size assigned to by ISHAPE in grey. The box with dashed red lines defines the colour and magnitude cuts. Only objects brighter than g=23 and with optical colours in the range $0.5<(g-z)<2$ were maintained in the sample. The final sample of GCs after visual inspection and size cut (see text for details) are indicated as blue squares.  
\textit{Right panel: }Optical/NIR ($(g-k)\,vs.\,g$) colour magnitude diagrams for the same objects as in the left panel. Objects inside the box with dashed lines that are grey dots in the left panel are shown as red triangles in the right panel. Blue squares are as in the left panel.}
\label{cmd}
\end{figure*}

In the left panel of Figure \ref{cmd} we show the optical colour magnitude diagrams (CMDs) for the GC candidates detected in all three bands and that had sizes successfully measured, ie., where the KING30 profiles were successfully convolved with the empirical PSF. Bimodality is readily seen for several galaxies (NGC\,4486, NGC\,4649, NGC\,4526 and NGC\,3377). 
The NGC\,4649 CMD looks different than the CMDs for the rest of the galaxies due to several faint, blue objects with $(g-z)<0.5$, which are likely to be contaminants from a companion spiral galaxy that can be seen in the upper right corner of the NGC\,4649 image in Figure \ref{images}.
In the right panel the optical/NIR CMDs for the same objects as in the left panel are shown.

\begin{figure}
\resizebox{\hsize}{!}{\includegraphics{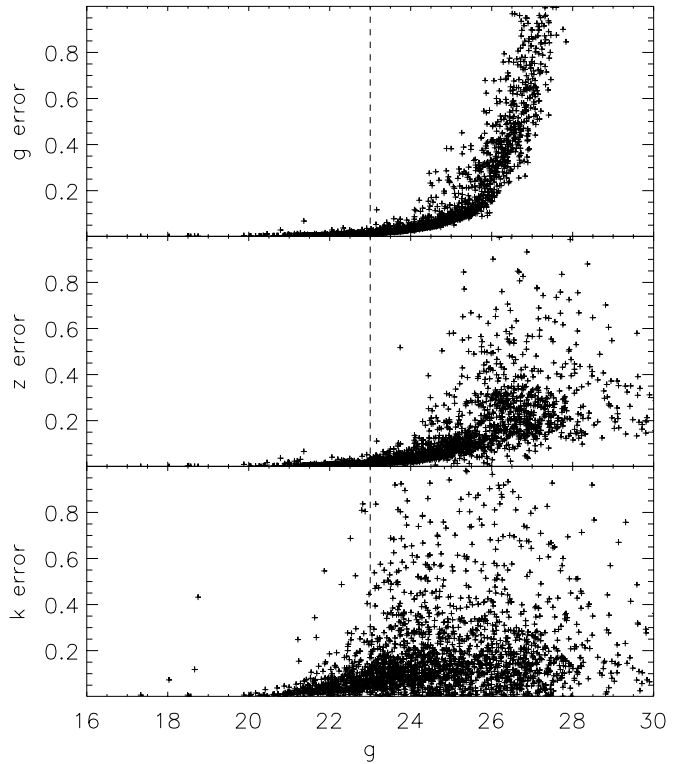}}
\caption[]{PHOT errors in g, z and K as a function of g magnitude for NGC\,4486 as an example. The dashed line defines the cut, only objects brighter than g=23 were kept in the sample.}
\label{err}
\end{figure}

\begin{figure}
\includegraphics[width=3.5in]{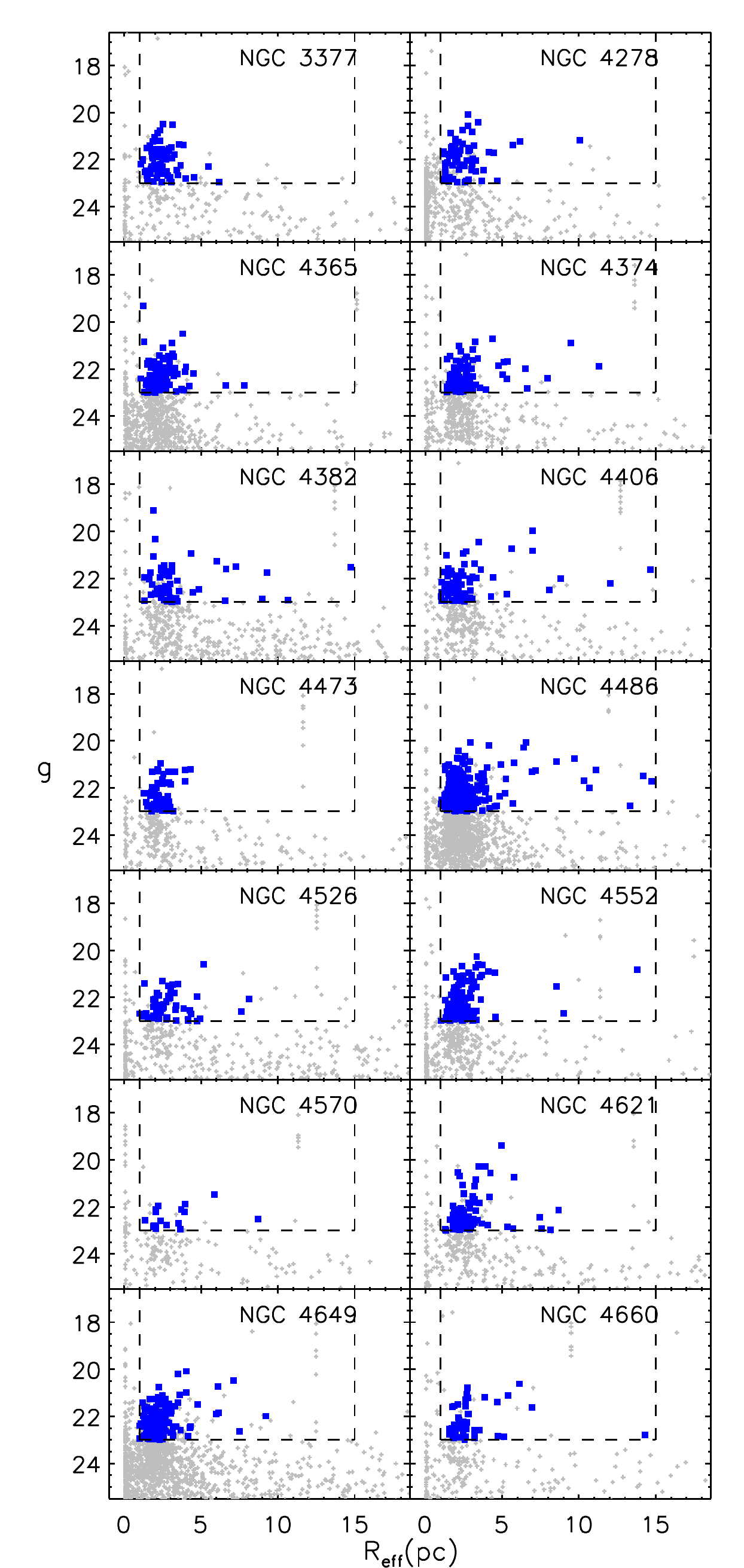}
\caption[]{The $g$ magnitude as a function of $\rm R_{eff}$. The box with dashed lines defines the size cut: $1<R_{\rm eff}(pc)<15$ and the magnitude cut $g<23$. The final sample, after visual inspection is indicated as blue squares, as in Fig. \ref{cmd}.}
\label{size_g}
\end{figure}

\begin{figure}
\includegraphics[width=7cm,angle=90]{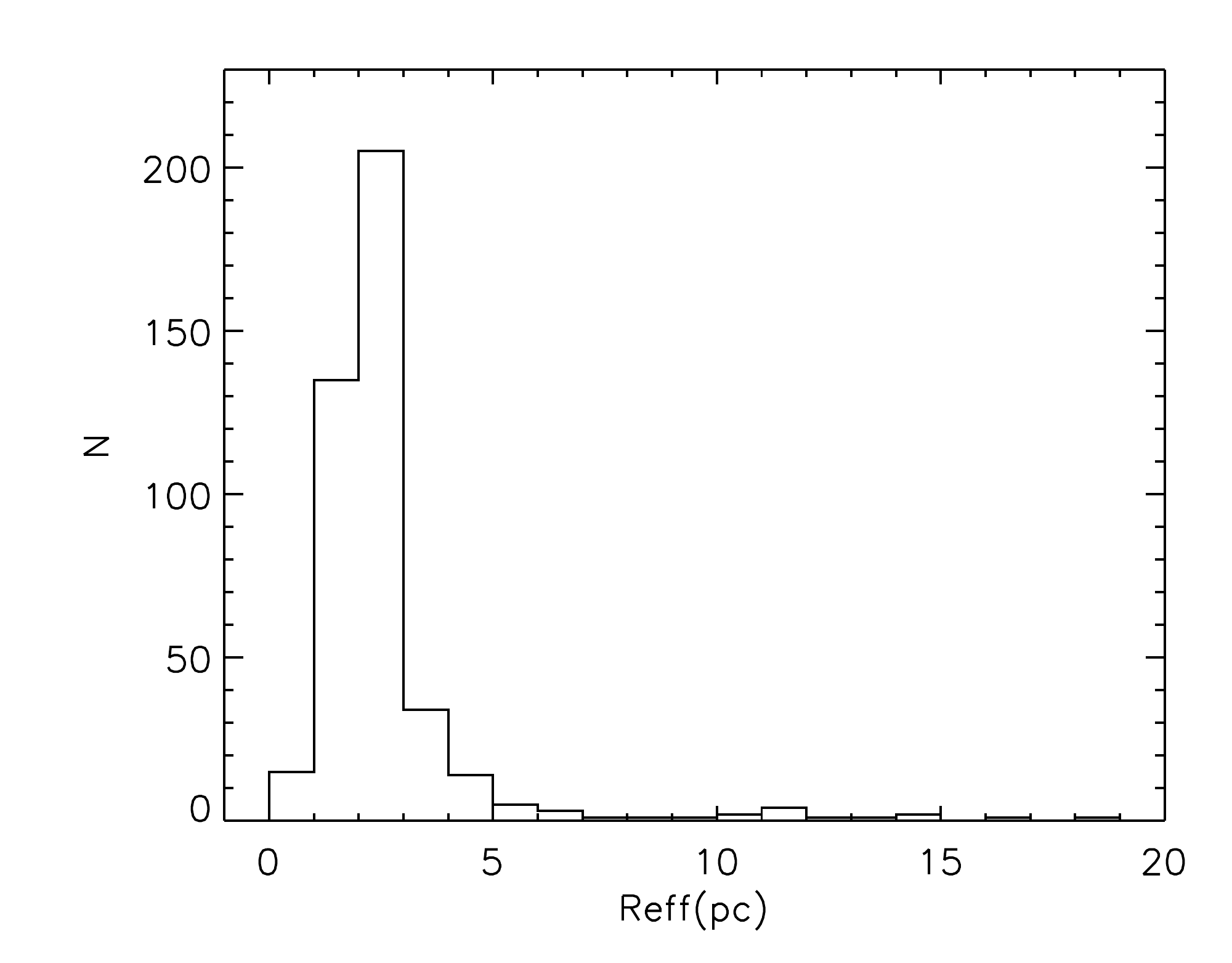}
\caption[]{The distribution of sizes for the GC candidates of NGC\,4486 as an ilustration after the colour and magnitude criteria were applied: $0.5<(g-z)<2.0$ and $g<23$.}
\label{sizecriteria}
\end{figure}

To obtain a cleaner sample of star clusters, only objects brighter than g=23 and with optical colours in the range $0.5<(g-z)<2$, expected for clusters in early-type galaxies (\citealt{peng06}), were kept in the sample. A box with dashed lines is shown in the left panel of Figure \ref{cmd} defining the regions that fall within these colour and magnitude cuts. The objects that fall inside the box in the left panel are indicated by a different colour/symbol in the right panel. 
We define a lower magnitude cut in an optical colour due to the larger errors in $K$, specially at fainter magnitudes.
The magnitude cut translates roughly into an error cut, as can be seen in Figure \ref{err}. 
This Figure shows the PHOT errors for $g$, $z$ and $K$.
Objects brighter than g=23 have g and z errors of about 0.05 mag while the bulk of objects have K-band errors below 0.2 mag. 
Following the ISHAPE measurements, we added a size range criterion. The objects had to have sizes between $1<R_{\rm eff}(pc)<15$ to be kept in the final sample of GCs. 
This is the range that includes the great majority of Milky Way GCs. 
The lower cut in sizes of $R_{\rm eff} (pc) >1$ removes objects that are essentially unresolved and foreground stars whereas the upper cut of $R_{\rm eff}(pc)=15$ excludes background galaxies. With these cuts one should keep essentially star clusters. Although some very distant background galaxies might still be contaminating the sample, we did not estimate the contamination from these. To illustrate the size criteria, we show the distribution of effective radii ($R_{\rm eff}$) for the NGC\,4486 GC candidates in Figure \ref{sizecriteria} that had already satisfied the colour and magnitude criteria: $0.5<(g-z)<2.0$ and $g<23$. Note the peak around $R_{\rm eff} \sim 2.5$, which is expected and is in accordance with \cite{jordan05}. By adopting $1\,<\,R_{\rm eff}\,<\,15$pc as criterion $\sim8\%$ of the candidates that had sizes successfully measured for NGC\,4486 drop out of the sample.
In Figure \ref{size_g} the $g$ magnitudes are shown as function of $\rm R_{eff}$ for the GC candidates that were detected in the three bands and that had sizes successfully measured, as in Fig.\,\ref{cmd}. The final sample of GCs is indicated by a different colour/symbol. Most GCs in the final sample have $\rm R_{eff}\sim 3$. Note the clear separation between point-like sources (stars) and GC candidates, $\lesssim1$\,pc.  
Moreover, we excluded all objects in NGC\,4649 that were located inside a rectangle of size $40\arcsec \, \times \,90\arcsec$ that contains the smaller companion galaxy.
 
\begin{figure}
\resizebox{\hsize}{!}{\includegraphics[width=43mm]{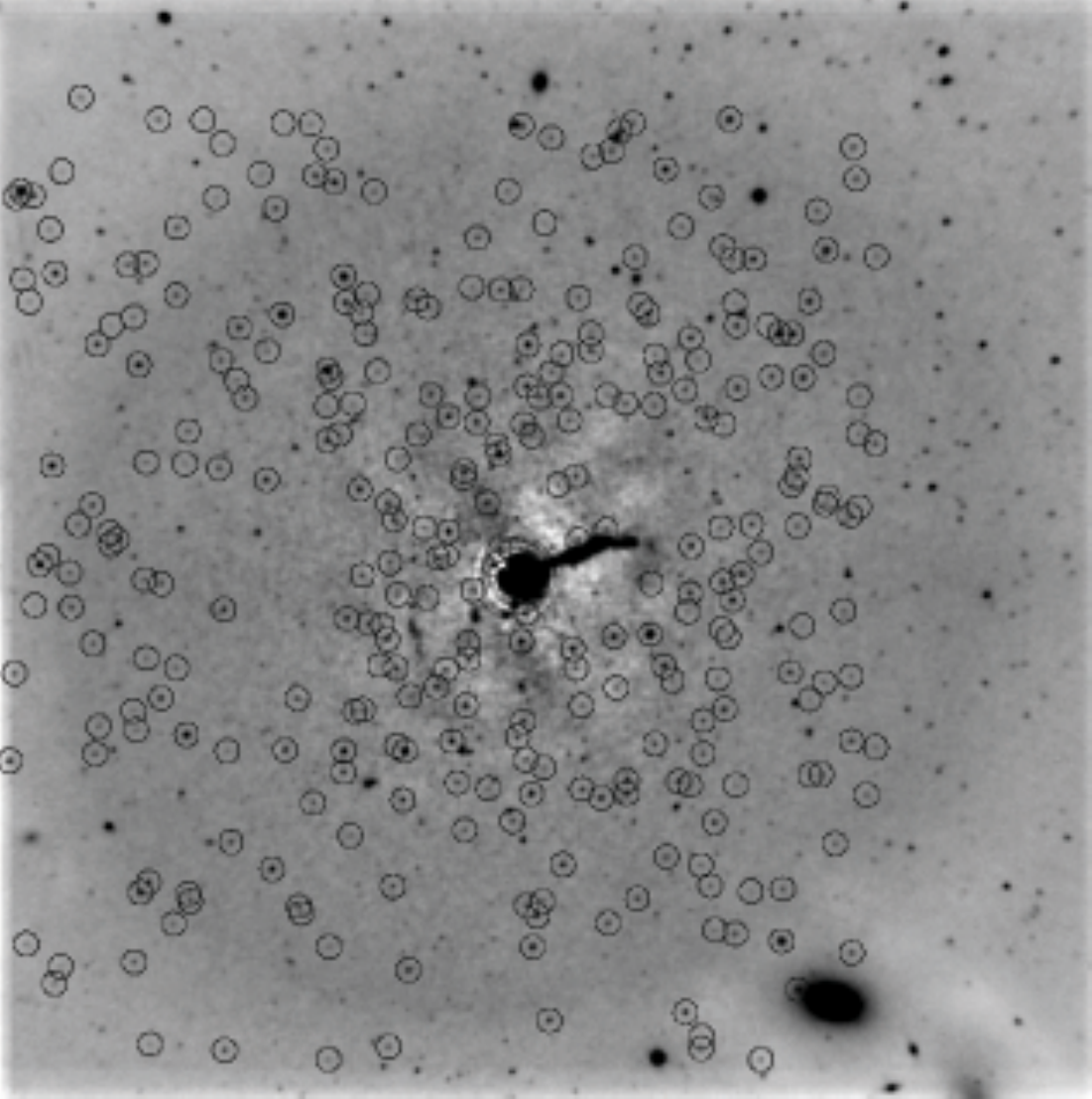}}
\caption[]{Final reduced and model-subtracted NGC\,4486 image with its final optical/NIR GCs as described above. North is up and east is left; the image scale is inverted.}
\label{m87_gimp}
\end{figure}

Furthermore, since the resolution of ACS is much superior to that of LIRIS, a final criterion for GC selection was added. If GC candidates are too close to each other in the ACS images they have a chance of appearing as one bigger source in the LIRIS K-band image, i.e they might be blended with another source in the K-band. The better the seeing, the smaller this chance. To take this into account the cluster candidates were visually inspected and the ones that looked like they might be blended with another source were flagged. 
We defined a minimum distance between neighbouring sources not to be blended with each other as $R_{\rm{blend}}$. We also assigned a criterion for the brightness of the neighbouring object. 
The neighbouring source should be at a distance $\le R_{\rm{blend}}$ and the flux of this neighbour should be $> 10\%$ of the flux of the GC candidate for it to be considered a blend.
For example, for a GC candidate of NGC\,4486 that was flagged as a possible blend, the separation between the neighbour and the GC candidate should be $\le$ 30 pix in the ACS image ($\sim1.5\arcsec$) and the flux of this neighbour should be $>$10\% of the flux of the GC candidate for it to be removed from the sample. $R_{\rm{blend}}$ was determined by visual inspection of the smallest distance between two sources in ACS to appear as two sources in LIRIS. $R_{\rm{blend}}$ is related to the seeing: if seeing  $=0.8 - 1.0 \arcsec, R_{\rm{blend}} = 25$\,$ACS_{\rm{pix}}$; if seeing $=1 - 1.2 \arcsec$, $R_{\rm{blend}}= 30$\,$ACS_{pix}$; if seeing $=1.3 - 1.5 \arcsec$, $R_{\rm{blend}} = 35$\,$ACS_{\rm{pix}}$. $R_{\rm{blend}}$ and the number of blended sources ($\rm N_{blend}$) for each galaxy are listed in Table \ref{jobsliris}. Together with eliminating the blends, we checked and excluded those GC candidates that were obvious background galaxies. Also, objects that fell in the residuals of the K-band model subtraction in the very centre ($\sim24\arcsec$) of the galaxy where removed.
The fraction of eliminated objects in the visual inspection ranged from $3\,-\,20\%$, and this depends on how cluster rich a galaxy is. For the majority of the cases, $10\%$ of the cluster candidates were eliminated; as expected, the richest cluster galaxy NGC\,4486 had $20\%$ of the candidates removed from the sample due to crowding. 
In Fig. \ref{m87_gimp} we show the final reduced and model subtracted NGC\,4486 image where we ran photometry on. GCs that fullfill the criteria defined above are shown as circles. 
In Table \ref{sampletab} the g, z and K photometry, astrometry and size information is given for a sample of these NGC\,4486 GCs shown in Figure \ref{m87_gimp}.

In summary, the selection criteria are:
\begin{itemize}
\item magnitude cut $g<23$.
\item optical colour cut  $0.5<(g-z)<2$.
\item size cut  $1<R_{\rm eff}<15$ eliminating background galaxies and foreground stars.
\item magnitude uncertainty cut $K_{\rm err}<0.5$.
\item visual inspection, flagging blends (see text for details) and still left over spurious objects such as obvious background galaxies and objects over the residuals of the K-band model subtraction in the centre of the galaxy.
\end{itemize}

\begin{table*}
\begin{scriptsize}
\centering
\begin{tabular}{ccccccccccc}
\hline\hline
Galaxy & ID & RA(J2000)  & DEC(J2000) &  $g$  & $\sigma g$   & $z$ & $\sigma z$  & $K$ & $\sigma K$& $R_{\rm{eff}}$(pc) \\
\hline
...&&&     &            &      &      &      &          &      &            \\  
NGC\,4486&1& 187.71700  &     12.36329 &      22.944     &   0.014   &    21.732    &    0.012  &       19.462     &   0.094     &          2.994   \\    
NGC\,4486&2& 187.69138  &     12.36336 &      22.572     &   0.010   &    21.242    &    0.008  &       19.369     &   0.059     &          2.599   \\     
NGC\,4486&3& 187.72313  &     12.36383 &      22.384     &   0.010   &    20.935    &    0.007  &       18.662     &   0.025     &          2.373   \\     
NGC\,4486&4& 187.69490  &     12.36397 &      22.560     &   0.011   &    21.028    &    0.007  &       18.697     &   0.023     &          2.316   \\     
NGC\,4486&5& 187.71342  &     12.36401 &      22.947     &   0.013   &    21.436    &    0.010  &       19.095     &   0.053     &          1.412   \\  
...&&&     &            &      &      &      &          &      &            \\  
\hline
\end{tabular}
\caption{ACS and LIRIS photometry and sizes for the GC system of NGC\,4486. The full version of this table, containing the other 13 GC systems presented in this study is available online at the CDS or upon request from the authors.}
\label{sampletab}
\end {scriptsize}
\end{table*}

\section{Summary}
In this paper, we have introduced a homogeneous survey of globular cluster systems in early-type galaxies from a combination of NIR and optical photometry. We have acquired deep K-band images for 14 early-type galaxies with the LIRIS instrument on the William Herschel Telescope and combined with the HST/ACS optical g (F475W) and z (F850LP) bands. Based on sizes, colours and optical brightness we have defined a sample of GCs for each galaxy. 
We have described the observing strategy and the data reduction procedures.
Comparing the LIRIS data to NTT/SOFI data (\citealt{lbs05}) for the galaxy NGC\,4365 we find a difference of $\sim\,0.1$\,mag showing that our photometry is reliable and agrees better with the ISAAC results by \cite{puzia02}.
The errors from PHOT for the GCs are found to be underestimated by roughly a factor of two.  

This survey has the great advantage of being homogeneous: data from the same instrument, with roughly the same exposure time, reduced and analysed in the same way.
It is also aided by homogeneous optical ACS/HST data, which helps the decontamination of the GC sample by its resolution, which allows size measurements.
In total, we have assembled a sample of $\sim$ 1300 GCs in 14 galaxies. For the majority of the galaxies we detect $\sim$ 70 GCs with $g\,\le\,23$. NGC\,4486 and NGC\,4649, the cluster-richest galaxies in the sample contain 301 and 167 GCs, respectively. These are the largest number of GCs so far for which NIR photometry has been obtained.
 
In the upcoming papers of the series we will address the science results of the data presented here. 

\begin{acknowledgements}
We thank Jos\'e Acosta Pulido for providing us with the IRAF package LIRISDR and Christoph Keller for comments that improved the paper.
ALCS acknowledges Michele Cantiello for useful discussions on photometric uncertainties.
\end{acknowledgements}


\begin{thebibliography}{}

	
\bibitem[Acosta Pulido et al.(2003)]{apulido03}Acosta Pulido, J. A., Ballesteros, E., Barreto, M., et al., 2003, ING Newsl., 7, 15 
\bibitem[Ashman \& Zepf(1992)]{az92}Ashman, K.M., Zepf, S,E., 1992, ApJ, 384, 50
\bibitem[Bedin et al.(2004)]{bedin04}Bedin, L.R., Piotto, G.P., Anderson, J. et al., 2004, ApJ, 605, 125
\bibitem[Beasley et al.(2002)]{beasley02} Beasley, M.A., Baugh, C. M., Forbes, D. A., Sharples, R. M., Frenk, C. S. , 2002, MNRAS, 333, 383
\bibitem[Bender(1998)]{bender98}Bender, R., 1988, A\&A, 202, 5
\bibitem[Bertin \& Arnouts(1996)]{bertin96} Bertin, E. \& Arnouts, S., 1996, A\&AS, 117, 393
\bibitem[Bridges et al.(2006)]{bridges06}Bridges, T., Gebhardt, K., Sharples, R. et al., 2006, MNRAS, 373, 157
\bibitem[Brodie et al.(2005)]{brodie05}Brodie J.P., Strader, J., Denicol\'o, G. et al., 2005, AJ, 129, 2643
\bibitem[Brodie \& Strader(2006)]{bs06} Brodie J.P. \& Strader J. 2006, ARA\&A, 44, 193
\bibitem[Cenarro et al.(2007)]{cenarro07} Cenarro,A.J., Beasley,M.A.,Strader,J., Brodie, J.P \& Forbes, D.A. 2007, AJ,134,391
\bibitem[Cantiello \& Blakeslee(2007)]{cb07} Cantiello, M. \& Blakeslee, J.P., 2007, ApJ, 669, 982
\bibitem[Cappellari et al.(2007)]{cap07}Cappellari, M., Emsellem, E., Bacon, R. et. al. 2007, MNRAS, 379, 418
\bibitem[C\^ot\'e et al.(1998)]{cote98} C\^ot\'e, P., Marzke, R.O., West, M.J., 1998, ApJ, 501, 554
\bibitem[C\^ot\'e et al.(2004)]{cote04} C\^ot\'e, P., Blakeslee, J.P., Ferrarese, L. et al., 2004, ApJS, 153, 223 
\bibitem[Cohen et al.(1998)]{cohen98} Cohen, J. G., Blakeslee, J. P. \& Ryzhov, A., 1998, ApJ, 496, 808
\bibitem[Cohen et al.(2003)]{cohen03} Cohen, J. G., Blakeslee, J. P. \& C\^ot\'e, 2003, ApJ, 592, 866
\bibitem[Chiosi \& Carraro(2002)]{cc02}Chiosi, C. \& Carraro, G.,  2002, MNRAS, 335, 335
\bibitem[Davies et al.(2001)]{davies01}Davies, R.L., Kuntschner, H., Emsellem, E. et al. 2001, ApJ, 548, 33
\bibitem[Elson \& Santiago(1996)]{es96}Elson, R. A. W. \& Santiago, B. X., 1996, MNRAS, 278, 617
\bibitem[Emsellem et al.(2007)]{emsellem07}Emsellem, E., Cappellari, M., Krajnovi\'c, D., et al., 2007, MNRAS, 379, 401
\bibitem[Ferrarese et al.(2006)]{ferrarese06}Ferrarese, L.,  C\^ot\'e, P.,Jord\'an, A. et al., 2006, ApJS, 164, 334
\bibitem[Forbes(1996)]{forbes96a}Forbes, D. A., 1996, AJ, 112, 1409
\bibitem[Forbes et al.(1996)]{forbes96}Forbes, D. A., Franx, M., Illingworth, G. D. \& Carollo, C. M., 1996, ApJ, 467, 126
\bibitem[Forbes et al.(1997)]{fbg97}Forbes, D.A., Brodie, J. P. \& Grillmair, C. J., 1997, AJ, 113, 1652
\bibitem[Forbes et al.(2004)]{forbes04}Forbes, D. A., Faifer, F, R., Forte, C. J. et al., 2004, MNRAS, 355, 608
\bibitem[Gebhardt et al.(2003)]{gebhardt03}Gebhardt, K., Richstone, D., Tremaine, S. et al., 2003, ApJ, 583, 92
\bibitem[G\'omez \& Richtler(2004)]{gr04}G\'omez, M. \& Richtler, T., 2004, A\&A, 415, 499
\bibitem[Goudfrooij et al.(2001)]{goud01b}Goudfrooij, P., Alonso, M. V., Maraston, C. \& Minniti, D., 2001, MNRAS, 328, 237
\bibitem[Goudfrooij et al.(2007)]{goud07}Goudfrooij, P., Schweizer, F., Gilmore, D., \& Whitmore, B., 2007, AJ, 133, 2737
\bibitem[Harris(2006)]{harris06} Harris, W. E., Whitmore, B. C.; Karakla, D. et al.,2006, ApJ, 636, 90
\bibitem[Harris(2009)]{harris09}Harris, W.E., 2009, ApJ, 703, 939
\bibitem[Hempel et al.(2007a)]{hempel07}Hempel, M., Zepf, S., Kundu, A., Geisler, D., \& Maccarone, T. J., 2007, ApJ, 661, 768
\bibitem[Hempel et al.(2007b)]{hempel07AA}Hempel, M., Kissler-Patig, M., Puzia, T. H., \& Hilker, M., 2007, A\&A, 463, 493
\bibitem[Hwang et al.(2008)]{hwang08}Hwang, Ho S., Lee, M. G.,  Park, H. S. et al., 2008, ApJ, 674, 869
\bibitem[Jord\'an et al.(2004)]{jordan04} Jord\'an, A., Blakeslee, J. P., Peng, E. W. et al., 2004, ApJS, 154, 509
\bibitem[Jord\'an et al.(2005)]{jordan05} Jord\'an, A., C\^ot\'e, P., Blakeslee, J. P. et al, 2005, ApJ, 634, 1002 
\bibitem[Jord\'an et al.(2009)]{jordan09} Jord\'an, A., Peng, E., Blakeslee, J. P. et al, 2009, ApJS, 180, 54
\bibitem[Kauffman et al.(1993)]{kauf93}Kauffmann, G., White, S. D. M. and Guiderdoni, B.,1993, MNRAS, 264, 201
\bibitem[Komatsu et al.(2009)]{kom09}Komatsu, E.; Dunkley, J.; Nolta, M. R. et al., 2009, ApJS, 180, 330
\bibitem[Kissler-Patig et al.(2002)]{kpbm02} Kissler-Patig, M., Brodie, J. P. \& Minniti, D., 2002, A\&A, 391, 441
\bibitem[Kotulla et al.(2008)]{kotulla08}Kotulla, R., Fritze, U. \& Anders, P., 2008, MNRAS, 387, 1149
\bibitem[Kormendy et al.(1998)]{kormendy98}Kormendy, J., Bender, R., Evans, A. S. \& Richstone, D., 1998, AJ, 115, 1823
\bibitem[Kundu \& Whitmore(2001)]{kw01}Kundu, A., Whitmore, B., 2001, AJ, 121, 2950
\bibitem[Kundu et al.(2005)] {kundu05}Kundu, A., Zepf, S. E., Hempel, M., et al., 2005, ApJ, 634, 41
\bibitem[Kundu \& Zepf(2007)]{kz07}Kundu, A., Zepf, S. E., 2007, ApJ, 660, 109
\bibitem[Kuntschner et al.(2010)]{kunt10}Kuntschner, H., Emsellem, E., Bacon, R., et al., 2010, \textit{MNRAS, accepted}
\bibitem[Lada \& Lada(2003)]{ll03}Lada \& Lada, 2003, ARA\&A, 41, 57
\bibitem[Larsen(1999)]{larsen99}Larsen,S.S.,1999, A\&AS, 139, 393
\bibitem[Larsen \& Richtler(1999)]{lr99}Larsen, S. S. and Richtler, T.,1999, A\&A, 345, 59
\bibitem[Larsen et al.(2001)]{larsen01}Larsen,S.S, Brodie, J.P., Huchra, J.P., Forbes, D.A.,Grillmair, C.J., 2001, AJ, 121, 2974
\bibitem[Larsen et al.(2003)]{larsen03}Larsen, S.S., Brodie,J.P., Beasley,M.A., et al., 2003, ApJ, 585, 767
\bibitem[Larsen et al.(2005)]{lbs05}Larsen, S. S., Brodie, J. P., Strader, J., 2005, A\&A, 443, 413
\bibitem[Lee et al.(2008)]{lee08} Lee, M. G., Park, H. S., Kim, E. et al., 2008, ApJ, 682, 135
\bibitem[Mackey(2008)]{mackey08} Mackey, A. D., Broby Nielsen, P., Ferguson, A. M. N., Richardson, J. C., 2008, ApJ, 681, 17
\bibitem[Manchado et al.(2004)]{manchado04}Manchado, A., Barreto, M., Acosta-Pulido et al., 2004, SPIE, 5492, 1094
\bibitem[Maraston(2005)]{m05}Maraston, C., 2005, MNRAS, 362, 799
\bibitem[Marigo et al.(2008)]{marigo08}Marigo, P., Girardi, L., Bressan, A. et al., 2008, A\&A, 482, 883
\bibitem[Mieske et al.(2006)]{mieske06}Mieske, S., Jord\'an, A., C\^ot\'e,P. et al., 2006, ApJ, 653, 193
\bibitem[Muratov \& Gnedin(2010)]{muratov10}Muratov, A. L. \& Gnedin, O. Y., 2010, ApJ, 718, 1266
\bibitem[Peng et al.(2006)]{peng06}Peng, E. W., Jord\'an, A. C\^ot\'e, P. et al., 2006, ApJ, 639, 95
\bibitem[Peng et al.(2008)]{peng08}Peng, E. W., Jord\'an, A., C\^ot\'e, P. et al., 2008, ApJ, 681, 197
\bibitem[Peng et al. (2009)]{peng09} Peng, E. W.; Jord\'an, A., Blakeslee, J. P. et al., 2009, ApJ, 703, 42
\bibitem[Percival et al.(2009)]{percival09}Percival, S. M., Salaris, M., Cassisi, S., Pietrinferni, A., 2009, ApJ, 690, 427
\bibitem[Piotto et al.(2007)]{piotto07}Piotto, G., Bedin, L. R.; Anderson, J.; et al. 2007, ApJ, 661, 53
\bibitem[Pierce et al.(2006)]{pierce06}Pierce, M., Bridges, T., Forbes, D. A. et al., 2006, MNRAS, 368, 325
\bibitem[Puzia et al.(2002)]{puzia02}Puzia, T. H., Zepf, S. E., Kissler-Patig, M., Hilker, M., Minniti, D. \& Goudfrooij, P.,2002, A\&A, 391, 453
\bibitem[Puzia et al.(2005b)]{puzia05}Puzia, T.H., Kissler-Patig, M., Thomas, D., et al. 2005, A\&A, 439, 997
\bibitem[Puzia et al.(2005a)]{puzia05a}Puzia, T. H., Perrett, K. M. \& Bridges, T. J., 2005, A\&A, 434, 909
\bibitem[Whitmore \& Schweizer(1995)]{ws95} Whitmore, B. C. \& Schweizer, F., 1995, AJ, 109, 960
\bibitem[Woodley et al.(2009)]	{woodley09} Woodley, K. A., Harris, W. E., Puzia, T. H. et al., 2009arXiv0911.0955W
\bibitem[Renzini(2006)]{renzini06}Renzini, A., 2006 ARA\&A, 44, 141
\bibitem[Rhode \& Zepf(2004)]{rz04}	Rhode, K. L. \& Zepf, S. E., 2004, AJ, 127, 302
\bibitem[Rhode et al.(2005)]{rhode05}Rhode, K.L., Zepf, S.E., Santos,R., 2005, ApJ, 630, 21
\bibitem[S\'anchez-Bl\'azquez et al.(2006)]{sb06}S\'anchez-Bl\'azquez P., Gorgas, J., Cardiel, N. \&Gonz‡lez, J. J., 2006, A\&A, 457, 809
\bibitem[Santiago et al.(2002)]{santiago02} Santiago, B., Kerber, L., Castro, R., de Grijs, R., 2002, MNRAS, 336,139
\bibitem[Schirmer(2007)]{cookbook}Schirmer, M., The Liris/WHT Cookbook
\bibitem[Schlegel et al.(1998)]{schlegel98}Schlegel, D. J., Finkbeiner, D. P. \& Davis, M., 1998, ApJ, 500, 525
\bibitem[Shapiro et al.(2010)]{shapiro10}Shapiro, K. L., Falc\'on-Barroso, J., van de Ven, G et al., 2010, MNRAS, 402, 2140
\bibitem[Sirianni et al(2005)]{sirianni05}Sirianni, M., Jee, M. J., Ben'tez, N. et al., 2005, PASP, 117, 1049
\bibitem[Strader et al.(2005)]{strader05}Strader,J., Brodie, J.P., Cenarro,A.J.,Beasley,M.A. \& Forbes, D.A. 2005, AJ,130,1315
\bibitem[Strader et al.(2006)]{strader06}Strader,J., Brodie, J.P., Spitler, L., Beasley, M. A., 2006, AJ, 132, 2333
\bibitem[Strader et al.(2007)]{strader07}Strader J., Beasley, M. A. \& Brodie, J. P., 2007, AJ, 133, 2015 
\bibitem[Tonry et al.(2001)]{tonry01}Tonry, J. L., Dressler, A., Blakeslee, J. P. et al., 2001,ApJ, 546, 681
\bibitem[Trager et al.(2008)]{trager08}Trager, S. C., Faber and S. M. Dressler, A., 2008, MNRAS, 386, 715
\bibitem[Tully et al.(1988)]{tully88}Tully, R. B., 1988, Sci, 242, 310
\bibitem[Yamada et al.(2006)] {yamada06}Yamada, Y., Arimoto, N., Vazdekis, A. \& Peletier, R. F., 2006, ApJ, 637, 200
\bibitem[Yoon et al.(2006)]{yoon06}Yoon, S.J. and Yi, S.K. and Lee, Y.-W., 2006, Sci, 311,1129
\bibitem[V\'eron-Cetty \& V\'eron(2006)]{veron06}V\'eron-Cetty, M. P., V\'eron, P., 2006, A\&A, 455, 773
\bibitem[de Zeeuw et al.(2002)]{zeeuw02}de Zeeuw, P. T., Bureau, M., Emsellem, E. et al., 2002, MNRAS, 329, 513

\end{thebibliography}
\end{document}